\pdfoutput=1
\documentclass[letterpaper,twocolumn,10pt]{article}
\usepackage{usenix-2020-09}

\usepackage{graphicx}
\usepackage{subcaption}
\usepackage{booktabs}
\usepackage{tabularx}
\PassOptionsToPackage{hyphens}{url}
\usepackage{url}

\usepackage{etoolbox}
\DeclareCaptionJustification{parfill}{\setlength{\parfillskip}{0pt plus 0.8\linewidth}}
\AtBeginDocument{\apptocmd{\thebibliography}{\setlength{\itemsep}{0pt plus 1pt}\setlength{\parsep}{0pt}\setlength{\parskip}{0pt}\raggedright\Urlmuskip=0mu\relax}{}{\typeout{WARNING: bibliography patch failed}}}
\newcommand{\sysname}{Ostinato}   %
\hypersetup{pdfsubject={}, pdfauthor={}, pdftitle={\sysname: Low-Latency CPU-GPU MoE Inference for Local Agents}}

\usepackage{amsmath}
\usepackage{amssymb}
\usepackage{mathtools}
\usepackage{amsthm}
\DeclareSymbolFont{largesymbols}{OMX}{cmex}{m}{n}
\usepackage{pifont}
\usepackage{colortbl}
\usepackage{placeins}
\usepackage{float}   %
\usepackage[capitalize,noabbrev]{cleveref}

\usepackage{xspace}
\newcommand{\sys}{{\upshape\textsc{\sysname}}\xspace}
\newcommand{\Ex}{\mathbb{E}}
\definecolor{ktc}{HTML}{8D99A6}\definecolor{layerc}{HTML}{6AAF6A}\definecolor{cachec}{HTML}{5B84B1}
\definecolor{ikc}{HTML}{C9789B}\definecolor{buunc}{HTML}{8E7CC3}\definecolor{ftc}{HTML}{D9625B}
\definecolor{oursc}{HTML}{F5B041}\definecolor{oursbg}{HTML}{FEF4E2}
\newcommand{\chip}[1]{\textcolor{#1}{\rule{0.85ex}{0.85ex}}\hspace{0.5em}}

\theoremstyle{plain}
\newtheorem{proposition}{Proposition}[section]
\usepackage{aliascnt}
\newaliascnt{lemma}{proposition}\aliascntresetthe{lemma}
\newaliascnt{observation}{proposition}\aliascntresetthe{observation}
\newaliascnt{corollary}{proposition}\aliascntresetthe{corollary}
\crefname{lemma}{Lemma}{Lemmas}\crefname{observation}{Observation}{Observations}\crefname{corollary}{Corollary}{Corollaries}

\begin{document}
\setlength{\parfillskip}{0pt plus 0.8\linewidth}
\raggedbottom  %
\pagestyle{plain}

\date{}

\newcommand{\figonecaption}{\sys achieves the lowest median tool-turn wait against default baselines and finishes sessions fastest in every group. Held-out SWE speedups over each group's slowest baseline (FreeToken: total time only, so the one-GPU group in (a) is relative to llama.cpp's cache); dashes: 512-token upload threshold; labels in (a): short-turn PCIe expert copies.}

\title{\Large \bf \sys: Low-Latency CPU--GPU MoE Inference for Local Agents}

\twocolumn[{%
  \vspace*{0.03in}%
  {\centering
    {\Large \bf \sys: Low-Latency CPU--GPU MoE Inference for Local Agents\par}%
    \vspace{0.13in}%
    \begin{tabular}{@{}c@{\hspace{3.5em}}c@{}}
      {\large\itshape Junning Zhu} & {\large\itshape Jianxiong Guo\textsuperscript{*}}\\[1pt]
      Beijing Normal-Hong Kong Baptist University & Beijing Normal University\\
      \texttt{t330025113@mail.bnbu.edu.cn} & \texttt{jianxiongguo@bnu.edu.cn}
    \end{tabular}\par}%
  \vspace{0.10in}%
  {\centering
    \includegraphics[width=\textwidth]{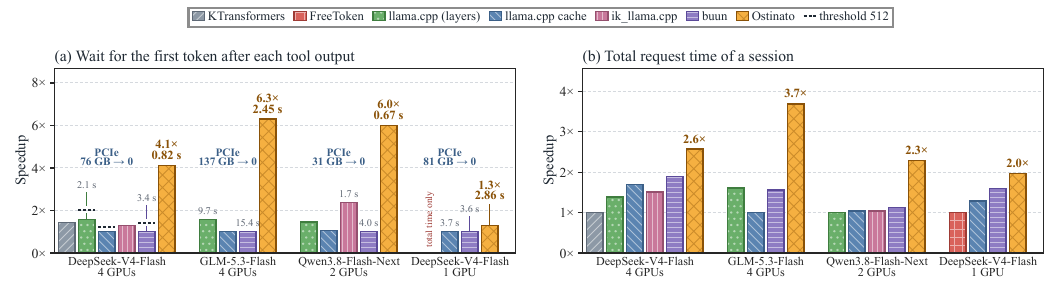}\par
    \captionof{figure}{\figonecaption}%
    \label{fig:teaser}\par}%
  \vspace{0.10in}%
}]
\thispagestyle{plain}

\begin{abstract}
Language-model agents increasingly run on local workstations with consumer GPUs, where quantized mixture-of-experts (MoE) models split routed experts between GPU and host memory. Prefix-cached agent sessions alternate single-token decode with short tool-return prefills, typically up to a few hundred tokens. Short tool turns touch more experts than consumer GPUs can hold, yet direct roughly 85\% of DeepSeek-V4-Flash's routes to a static GPU pool while leaving 3.3--5.5 tokens per cold expert. Widely used local engines such as llama.cpp and KTransformers either copy tens of gigabytes over PCIe per turn, compute every route on the CPU while idling resident GPU experts, or budget GPU memory by whole layers and uniform per-layer quotas. We present \sys, a CPU--GPU MoE engine that reuses one static cross-layer GPU pool in place across every forward pass. During decode and short tool turns, \sys computes resident routes on the GPUs and sparse cold routes on the CPU, so short turns copy no expert weights; on long prefills, it deduplicates cold uploads; and for multi-GPU decode, all-zero expert slots with device-side synchronization capture each GPU's decode step in one CUDA graph. Across held-out SWE and $\tau^2$-bench sessions on three MoE models, \sys cuts median tool-turn waits by 2.5--6.6$\times$ and finishes sessions 1.36--3.83$\times$ faster than default baselines, staying 1.30--1.94$\times$ faster when baselines align their upload threshold and weight loading. It finishes sessions fastest across seven agent workloads, one to four GPUs, and up to eight concurrent sessions, while its cross-layer pool speeds up 335\,GB DeepSeek-V4.1-Flash by 1.81--2.19$\times$ over llama.cpp's expert cache.
\end{abstract}

\section{Introduction}
\label{sec:intro}

{\renewcommand\thefootnote{}\footnotetext{\textsuperscript{*}Corresponding author.}}

Language-model agents that repair software repositories~\cite{yang2024sweagent,wang2025openhands} or serve customers through tools~\cite{barres2026tau2bench} increasingly run on workstations owned by their users~\cite{chen2025ktransformers,yang2026freetoken,ggml2026llamacpp}, keeping private code, records, and tool outputs on the local machine. These agents rely on open-weight mixture-of-experts (MoE) models~\cite{shazeer2017outrageously,fedus2022switch,deepseekai2026deepseekv4,yang2025qwen3,glm5team2026glm5} whose routed experts alone exceed consumer GPU memory: DeepSeek-V4-Flash holds 147\,GB of routed experts against roughly 75\,GB of GPU capacity left after dense weights and key--value (KV) state on four consumer GPUs. Because compressing experts or altering routing can change generated tool calls~\cite{zhou2025floe,zhou2025oraclemoe,zhu2026remoe,oncescu2026opportunistic}, local engines preserve the model's unmodified weights and routing functions and place remaining experts in host DRAM. Splitting experts across two memories gives each \emph{route}, one token's use of one routed expert, three execution paths. A GPU executes the route in place if the expert is \emph{resident}; otherwise the expert is \emph{cold}, and the engine either computes it on the CPU from host memory or copies it over PCIe to a GPU.

\begin{figure}[t]
  \centering
  \includegraphics{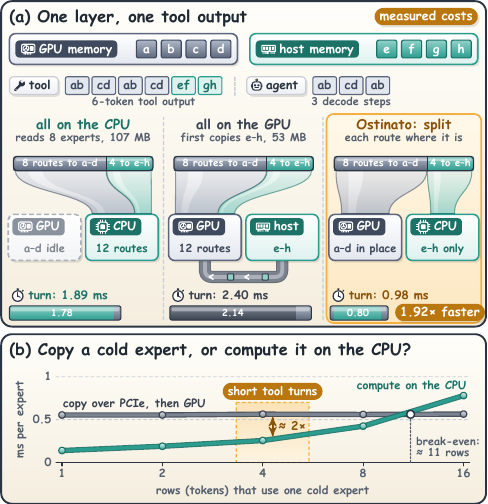}
  \caption{Reusing resident experts in place beats any single path. (a)~Single-layer example parameterized by cold-read per-expert costs on DeepSeek-V4-Flash (\cref{sec:background-insights}); teal: CPU, slate: GPU, charcoal: PCIe copies. (b)~Measured per-expert latency of PCIe copy versus CPU execution (amortized across 12 active experts on 32 threads without cache flushing between runs); shaded band: short tool turns.}
  \label{fig:toy}
\end{figure}

In prefix-cached agent sessions, each returning turn reuses prior KV states and prefills only the new tokens, so short tool turns of up to a few hundred tokens alternate with single-token decode steps. Short prefills touch more distinct experts than the GPUs can hold, yet most of their token--expert selections land on experts already resident there. On DeepSeek-V4-Flash, a short turn of 128--511 tokens touches 51--74\% of all experts (93.1\,GB on average, above the GPUs' expert capacity), while a static GPU pool serves about 85\% of its routes and leaves 3.3--5.5 tokens per cold expert, far below the PCIe copy break-even (\cref{fig:toy}b).

Mainstream local engines either upload the experts a short prefill misses or leave resident GPU experts idle, because they commit each prefill pass to a single path or budget GPU memory by layers. Expert caches in official llama.cpp~\cite{ggml2026llamacpp,am17an2026llama}, ik\_llama.cpp~\cite{ikawrakow2026ikllama}, and buun~\cite{spiritbuun2026cuda} keep popular experts on the GPUs for decode, yet route prefills through one path chosen by an \emph{upload threshold}: above it they copy tens of gigabytes over PCIe per turn, and below it they run every route on the CPU. Either choice waits 2.4--3.4\,s after each tool output on DeepSeek-V4-Flash coding sessions, versus 0.82\,s with \sys (\cref{fig:teaser}a). Whole-layer placement in llama.cpp runs offloaded layers on the CPU during every decode step, FreeToken~\cite{yang2026freetoken} streams whole layers in prefill, and KTransformers~\cite{chen2025ktransformers} via SGLang's~\cite{zheng2024sglang} \texttt{kt-kernel} uses uniform 58-expert per-layer placement and waits 2.36\,s per coding tool turn (\cref{sec:background-engines,sec:eval-main}).

Our central insight is that a local agent engine should reuse one static GPU-resident expert pool across \emph{every} forward pass, running resident routes in place and sparse cold routes on the CPU without letting weight copies or per-layer CPU--GPU joins consume the gain. Even in an eight-expert, four-slot single-layer example (\cref{fig:toy}a), splitting resident and cold routes finishes in 0.98\,ms versus at least 1.89\,ms for any single path (\cref{sec:background-insights}). On DeepSeek-V4-Flash, GPU-only short turns must transfer 17.7--81.0\,GB on average across four to one GPUs regardless of caching policy. Prior hybrid engines explored pieces of per-expert splitting: Fiddler~\cite{kamahori2025fiddler} places popular experts statically on one GPU and chooses CPU execution or upload per expert via eager PyTorch dispatch, and KTransformers~\cite{chen2025ktransformers} executes routed experts on the CPU inside a single-GPU CUDA graph using host callbacks. Realizing this split for multi-GPU local agents combines one per-GPU cross-layer pool across short prefills and decode with packed CPU short turns and device-side decode graphs.

\sys (\cref{fig:design}) pairs two placement-and-execution decisions with one runtime mechanism that realizes them at decode speed. First, \textbf{cross-layer expert residency} fills one shared pool per GPU by routing frequency; at matched GPU budgets, exact cross-layer pools leave 2.2--8.5$\times$ fewer cold decode routes than whole-layer placement. Second, \textbf{split execution} computes resident routes on the GPUs in place across every pass, runs sparse cold routes on a packed, overlapped CPU branch during decode and short tool turns, so short turns copy no expert weights, and switches once to deduplicated uploads on long prefills. Third, because per-layer splitting introduces a CPU--GPU join on every MoE layer, \textbf{a device-side decode graph per GPU} pads cold positions with all-zero expert slots (\emph{zero slots}) so multi-GPU decode replays as one CUDA graph per GPU and skips CPU wakeups on all-resident layers, while lending idle prefill buffers as \emph{borrowed memory} for recurring cold decode routes (\cref{sec:design-decode}).

We build \sys in llama.cpp and evaluate it across five MoE models up to 335\,GB on up to four RTX~4090D GPUs against KTransformers, FreeToken, official llama.cpp, ik\_llama.cpp, and buun. Across SWE and $\tau^2$-bench sessions on DeepSeek-V4-Flash, GLM-5.3-Flash~\cite{zai2026glm53flash}, and Qwen3.8-Flash-Next~\cite{qwen2026qwen38flashnext}, \sys cuts median tool-turn waits by 2.5--6.6$\times$, finishes sessions 1.36--3.83$\times$ faster than default baselines, and beats a \emph{stitched baseline floor} combining the fastest baseline prefill with the fastest baseline decode by 1.20--2.29$\times$. When baselines align their upload threshold to 512 tokens and match weight loading, \sys stays 1.30--1.94$\times$ faster and beats the aligned stitched floor by 1.17--1.69$\times$, while forcing short-turn uploads on the same static pool slows prefill by 2.10$\times$. \sys finishes sessions fastest across five additional agent workloads (spanning \texttt{mini-swe-agent}, Terminal-Bench~2.1~\cite{merrill2026terminalbench,buchanan2026benchmarks}, Claude Code~\cite{anthropic2026claudecode}, and BFCL~v4~\cite{patil2025bfcl}), one to four GPUs, and up to eight concurrent sessions, while its cross-layer pool accelerates the 335\,GB DeepSeek-V4.1-Flash~\cite{deepseekai2026deepseekv41flash} by 1.81--2.19$\times$ over llama.cpp's expert cache.

This paper makes three contributions:
\begin{itemize}
  \item \textbf{Workload characterization and GPU-only transfer bounds.} We characterize prefix-cached agent turns across three MoE models: wide expert footprints, concentrated static-pool routes, and sparse cold-expert reuse. We also establish policy-independent GPU-only transfer lower bounds averaging 17.7--81.0\,GB per short turn on DeepSeek-V4-Flash across four to one GPUs (\cref{sec:background,app:bounds}).
  \item \textbf{Resident-in-place execution and device-side multi-GPU decode.} We design and build \sys in llama.cpp, combining cross-layer per-GPU frequency pools, phase-adaptive split execution for short and long prefills, and zero-slot CUDA decode graphs with device-side synchronization across three MoE model families (\cref{sec:design,app:derivation}).
  \item \textbf{Controlled evaluation and generality.} \sys finishes sessions faster than default and aligned baselines across all main cells, isolates execution from residency through controlled ablations and a short-turn cost boundary, and generalizes across seven workloads, models up to 335\,GB, smaller resource budgets, and concurrent sessions (\cref{sec:eval,app:sessions,sec:eval-qwen}).
\end{itemize}

\section{Background and Motivation}
\label{sec:background}

\subsection{Three Paths for a Route}
\label{sec:background-moe}

An MoE layer~\cite{shazeer2017outrageously,lepikhin2021gshard} replaces the feed-forward network of a transformer block with a router, $E$ routed experts of which each token selects $k$, and shared experts that every token uses. A forward pass over $n$ tokens therefore issues $nk$ routes in each layer. Attention, router, and shared-expert weights reside in GPU memory alongside the routed experts that fit beside them, called \emph{resident} experts, while the remaining \emph{cold} experts reside in host memory.

The three execution paths for a route have asymmetric fixed and marginal costs. One MXFP4 expert of DeepSeek-V4-Flash occupies 13.4\,MB: reading it once from host memory at 87\,GB/s takes 0.154\,ms, and CPU execution adds 0.046\,ms per routed token. Copying it over PCIe~4.0 at 25\,GB/s takes 0.535\,ms regardless of token count, whereas a GPU computes a resident route in about 0.02\,ms in decode and roughly half that per route when batched in prefill. Measured across three layers of DeepSeek-V4-Flash, copying a cold expert over PCIe pays off only once roughly 11 tokens of the forward pass select it (\cref{fig:toy}b).

\subsection{Agent Turns: Two Phases and Three Facts}
\label{sec:background-agents}

\begin{figure*}[t]
  \centering
  \includegraphics[width=\textwidth]{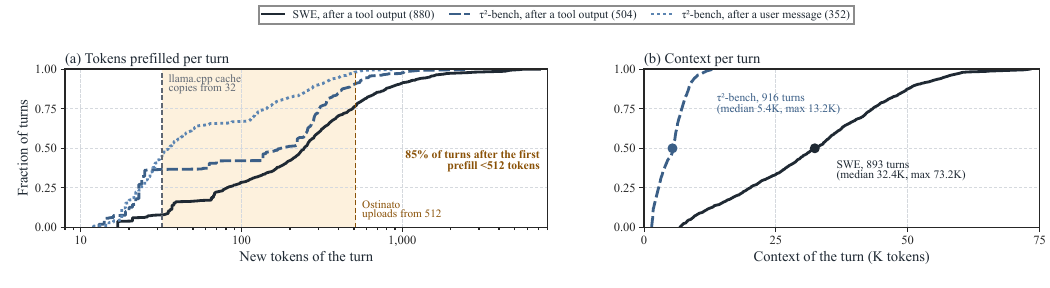}
  \caption{Held-out turns of DeepSeek-V4-Flash. (a)~New tokens prefilled per turn after the first, by turn type; dashed lines mark llama.cpp's 32-token default upload threshold and \sys's 512-token threshold. (b)~Prefix context extended by each turn.}
  \label{fig:workload}
\end{figure*}

An agent session alternates model generation with external tool calls~\cite{zhu2026tracelab,yuan2026agentic}. Each returning turn appends the latest tool output or user message, reuses the KV cache of prior turns, and prefills only the new tokens before decoding the next action. Across held-out sessions (42 to 100 turns on coding sessions), returning turns prefill a median of 245 new tokens on SWE and 63 on $\tau^2$-bench (248 and 118 across all turns; \cref{tab:workloads}), with 85.0\% of post-first turns prefilling fewer than 512 tokens while extending contexts up to 73.2K tokens (\cref{fig:workload}).

On short returning turns of 128 to 511 new tokens, three empirical facts hold simultaneously across agent sessions:
\begin{enumerate}
  \item \textbf{Broad weight footprint beyond GPU capacity.} A single short tool turn on DeepSeek-V4-Flash touches 51--74\% of all routed experts, or 93.1\,GB on average (up to 172\,GB on GLM-5.3-Flash, \cref{fig:footprint}), exceeding our four consumer GPUs' roughly 75\,GB expert budget.
  \item \textbf{Concentrated routes on the static GPU pool.} On native SWE development turns of 128--511 tokens, our static cross-layer pool covers 84.6\% of routes on DeepSeek-V4-Flash from 51.8\% of expert bytes (84.1--84.7\% across synthetic lengths $N \in \{128, 343, 511\}$) and 75.4\% on Qwen3.8-Flash-Next from 48.4\% of bytes, dropping to 50.3\% on flatter-routing GLM-5.3-Flash from 38.7\% of bytes. On DeepSeek-V4-Flash held-out decode steps, the pool serves 88.6--92.1\% of routes.
  \item \textbf{Sparse cold-expert reuse.} On DeepSeek-V4-Flash, the remaining cold routes spread across 46--61 distinct cold experts per layer, averaging only 3.3--5.5 tokens per touched cold expert, well below the roughly 11-token copy break-even (\cref{fig:toy}b).
\end{enumerate}

\subsection{Design Space of CPU--GPU MoE Engines}
\label{sec:background-engines}

\begin{table}[t]
  \caption{CPU--GPU MoE engine design choices.}
  \label{tab:designspace}
  \centering
  \footnotesize
  \setlength{\tabcolsep}{3.5pt}
  \begin{tabularx}{\columnwidth}{@{} >{\raggedright\arraybackslash}p{2.25cm} >{\raggedright\arraybackslash}p{1.45cm} >{\raggedright\arraybackslash}X >{\raggedright\arraybackslash}X @{}}
    \toprule
    Engine & Residency & Short prefill & Decode step \\
    \midrule
    llama.cpp (layers) & whole layers & CPU layers by threshold & per-segment graphs \\
    llama.cpp, ik\_llama.cpp, buun & dynamic expert cache & single path by threshold & per-segment graphs \\
    Fiddler~\cite{kamahori2025fiddler} & cross-layer popularity & per-expert CPU/upload & eager host loop (1 GPU) \\
    KTransformers~\cite{chen2025ktransformers} & 0 routed on GPU & AMX CPU prefill & 1 graph, host callbacks \\
    \texttt{kt-kernel} (SGLang) & default uniform quota & split or load layers & TP graph, host callbacks \\
    \rowcolor{oursbg}\sys & per-GPU cross-layer pool & split packed CPU branch & 1 device graph per GPU \\
    \bottomrule
  \end{tabularx}
\end{table}

\Cref{tab:designspace} summarizes how existing CPU--GPU engines choose expert residency, short-prefill execution, and decode synchronization. \textbf{Whole-layer placement}, official llama.cpp's default~\cite{ggml2026llamacpp}, keeps contiguous layers on GPUs, runs offloaded layers on the CPU in decode (47.4\,ms/token on DeepSeek-V4-Flash versus 31.4\,ms with llama.cpp's expert cache), and copies offloaded layers over PCIe on prefills at or above its upload threshold. \textbf{Expert caches} in official llama.cpp~\cite{am17an2026llama}, ik\_llama.cpp~\cite{ikawrakow2026ikllama,samueloliveirads2026add}, and buun~\cite{spiritbuun2026cuda} keep popular experts on the GPUs for decode, yet above buun's 2--7-token hybrid path for very short passes, every prefill follows one path set by an upload threshold: 32 tokens by default in llama.cpp and buun, and about 1,366 in ik\_llama.cpp on DeepSeek-V4-Flash. At or above this threshold, PCIe copies consume 86\% of wait in llama.cpp's cache and 78\% in buun. Short turns copy 76.5\,GB on DeepSeek-V4-Flash, 136.7\,GB on GLM-5.3-Flash, and 31.5\,GB on Qwen3.8-Flash-Next, with wait tracking $0.18\,\text{s} + \text{bytes}/24.2\,\text{GB/s} + 0.61\,\text{ms/token}$ ($R^2 = 0.9998$) on DeepSeek-V4-Flash.

Prior hybrid engines introduced parts of per-expert splitting in different regimes. \textbf{Fiddler}~\cite{kamahori2025fiddler} places popular experts across layers offline (56 of 256 on Mixtral-8x7B) and selects GPU, CPU, or upload per expert, yet runs single-GPU 16-bit eager PyTorch loops without branch overlap or CUDA graphs. \textbf{KTransformers}~\cite{chen2025ktransformers} runs routed experts on AMX CPUs inside a single-GPU CUDA graph via host callbacks; its SGLang \texttt{kt-kernel} backend adds TP=4 GPU residency, placing expert IDs 0--57 per layer under its default uniform strategy (optional cross-layer frequency placement is also supported) and waiting 2.36\,s per coding tool turn on DeepSeek-V4-Flash (\cref{sec:eval-main}).

\subsection{Why Single-Path Prefills and Better Caching Policies Stall}
\label{sec:background-example}
\label{sec:background-insights}

\begin{figure}[t]
  \centering
  \includegraphics{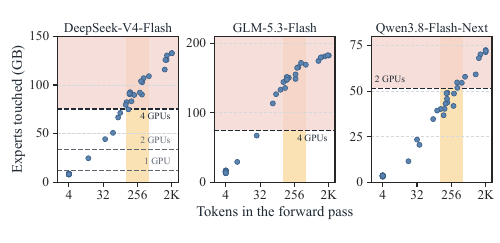}
  \caption{Expert bytes touched per prefill pass on a development session by pass length. Dashed lines: GPU expert budgets; red: bytes beyond budget that any GPU-only pass must copy; shaded band: short tool turns.}
  \label{fig:footprint}
\end{figure}

Running prefills through a single path leaves short tool turns stalled even in small synthetic settings or with an oracle replacement policy (\cref{app:bounds}). In \cref{fig:toy}a, one top-2 layer with eight 13.4\,MB experts a--h and four GPU slots (a--d) processes six tokens routed to ab, cd, ab, cd, ef, gh and three decode steps routed to ab, cd, ab. Under our workstation's measured per-expert costs for this fixed route sequence, a CPU-only prefill reads all eight experts in 1.78\,ms and a GPU-only prefill copies e--h in 2.14\,ms, so every single-path policy needs at least $1.78 + 0.11 = 1.89$\,ms. Running resident routes in place on GPU (0.075\,ms) and cold routes on CPU (0.80\,ms) finishes in 0.98\,ms even sequentially. At full scale, Observations~1 and~2 show why neither better GPU caching nor raising the upload threshold removes the stall.

\textbf{Observation 1: capacity-forced bytes under any GPU-only policy.} Short tool turns of 128 to 511 tokens exceed the GPUs' expert budget (\cref{fig:footprint}): 75.4\,GB on four-GPU DeepSeek-V4-Flash, 74.4\,GB on four-GPU GLM-5.3-Flash, and 37.6\,GB at two-GPU Qwen3.8-Flash-Next's configured cache (or 51.5\,GB if the cache used all free GPU memory; \cref{app:config} lists alternative cache sizes). Regardless of caching policy, any GPU-only pass over the same stored weights and routing must transfer at least the touched bytes beyond this budget (\cref{eq:weightcut}), averaging 17.7\,GB per development short turn on four-GPU DeepSeek-V4-Flash (59.4\,GB on two GPUs, 81.0\,GB on one GPU), 79.8\,GB on GLM-5.3-Flash, and 8.6\,GB on Qwen3.8-Flash-Next (0.5\,GB against all free GPU memory). Carrying cache state across alternating prefill and decode passes compounds this bound: across 4,000 DeepSeek-V4-Flash passes, even offline-optimal Belady/MIN caching with bypass loads 1,154.1\,GB, 27.4\% above the 905.9\,GB under free per-phase resets (\cref{app:bounds}).

\textbf{Observation 2: raising thresholds idles GPU residency on sparse cold routes.} Because each cold expert receives only 3.3--5.5 tokens in short tool turns, copying it costs 1.8--2.4$\times$ more than CPU execution (\cref{fig:toy}b). Sweeping buun's upload threshold on a development session improves prefill from 352.7\,s at 128 tokens to 310.3\,s at 512 before rising to 323.6\,s at 1,024. However, raising llama.cpp or buun to threshold 512 eliminates PCIe copies only by forcing the CPU to evaluate all six experts per token while leaving resident GPU experts idle, so coding agents still wait 2.37--2.78\,s versus 0.82\,s with \sys.

\subsection{Residency Granularity and Synchronization Overhead}
\label{sec:background-granularity}

\textbf{Observation 3: residency granularity governs decode cold routes.} Consider one GPU holding four equal-size slots across two top-2 MoE layers processing six tokens (12 routes per layer): Layer~1 routes to ab, ac, bc twice (frequencies $a=b=c=4$, $d=0$) and Layer~2 to wx, wy, wz twice ($w=6$, $x=y=z=2$). Optimal whole-layer placement keeps Layer~1 and leaves all 12 Layer~2 routes cold; a uniform two-slot quota per layer keeps $\{a,b\}$ and $\{w,x\}$, leaving eight cold routes; a cross-layer pool selects $\{w,a,b,c\}$ and leaves six. At scale on development requests under matched per-GPU bytes (\cref{tab:granularity}), an exact cross-layer pool cuts cold decode routes by 2.2--8.5$\times$ over whole-layer placement, while an optimal uniform model quota leaves 12.2--55.1\% more cold routes.

\textbf{Observation 4: splitting adds per-layer CPU--GPU joins in decode.} Serving resident routes on GPU and cold routes on CPU introduces a CPU--GPU join into every MoE layer (43 times per token on DeepSeek-V4-Flash, roughly 0.6\,ms per layer total). Eager host-side per-expert dispatch as in Fiddler triggers over 7,000 CUDA kernel launches per token, consuming 73\% of GPU time~\cite{chen2025ktransformers}. Even inside llama.cpp, running the two branches serially adds 4.9\,ms per step (\cref{app:derivation}), and per-segment CUDA graphs decode at 29.36\,ms/step, slower than buun's 28.65\,ms. Capturing one device-side graph per GPU drops decode latency to 27.43\,ms/step (\cref{sec:eval-residency}).

Observations~1--2 motivate phase-adaptive split execution and deduplicated upload (\cref{sec:design-prefill}), Observation~3 motivates per-GPU cross-layer residency (\cref{sec:design-model}), and Observation~4 makes device-side decode graph capture per GPU necessary to realize those route savings in decode (\cref{sec:design-decode}).

\section{Design and Implementation}
\label{sec:design}

\begin{figure*}[t]
  \centering
  \includegraphics[width=\textwidth]{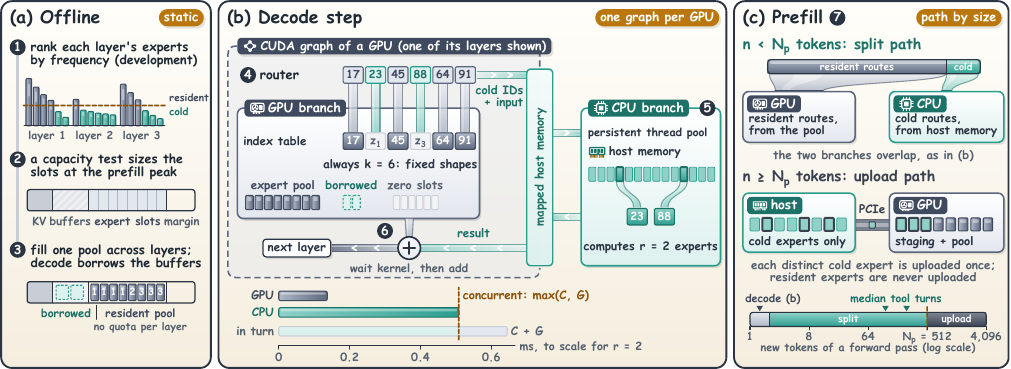}
  \caption{Overview of \sys: resident experts execute on the GPU in every forward pass while cold routes adapt by phase. (a)~Offline residency (\ding{182}--\ding{184}): one cross-layer cut fills each GPU's shared pool (digits: layer index of each resident expert), and decode borrows idle prefill buffers. (b)~One CUDA decode graph per GPU using zero slots for cold positions alongside a persistent CPU branch (\ding{185}--\ding{187}). (c)~Size-dependent prefill paths (\ding{188}), marking $\tau^2$-bench and SWE median tool turns.}
  \label{fig:design}
\end{figure*}

\subsection{System Overview}
\label{sec:design-overview}

\sys operates in two stages, offline residency planning and online execution, numbered as in \cref{fig:design}.

\textbf{Offline residency planning.} Run once before deployment, this stage selects which experts reside on each GPU (\cref{fig:design}a). \sys profiles development sessions and ranks every routed expert by routing frequency, step~\ding{182}. In step~\ding{183}, a capacity test determines each GPU's slot budget by requiring the longest development request, including full tool definitions, to complete while preserving a fixed free-memory margin on every GPU. In step~\ding{184}, \sys fills each GPU's pool with the highest-ranked experts across all layers assigned to that GPU, allocating separate slot groups when expert byte sizes differ across layers (\cref{sec:design-models}) and loading the static residency set at start-up.

\textbf{Online execution.} At start-up, \sys populates each GPU's expert pool and builds device and host index tables mapping every expert either to a pool slot or to the CPU. After the router selects $k$ experts in a layer, step~\ding{185} partitions routes into a \emph{GPU branch} over resident experts and a \emph{CPU branch} over cold experts. During decode, step~\ding{186} runs a persistent CPU thread pool on cold routes while the GPU computes resident routes, and step~\ding{187} synchronizes both branches at a device gather kernel before summing their outputs. On two or four GPUs, all layers on a GPU replay as a single CUDA graph while idle prefill buffers host temporary cold experts as \emph{borrowed memory} (\cref{sec:design-decode}). In step~\ding{188}, prefill passes choose between split execution and deduplicated upload by pass length (\cref{sec:design-prefill}), preserving the unmodified model's weights and routing functions in every pass.

\subsection{Cross-Layer Expert Residency}
\label{sec:design-model}

\textbf{Two branches and their costs.} In a decode step, MoE layer $\ell$ partitions its $k$ selected experts into two branches. Let $r_\ell$ count the layer's \emph{cold routes}, token--expert selections that target cold experts. Because each GPU reserves $k$ all-zero expert slots (\emph{zero slots}, \cref{sec:design-decode}) and maps every cold selection to its position's zero slot, the GPU branch runs fixed-shape kernels taking a constant time $G_\ell$; on the device-side decode graph (\cref{sec:design-decode}), all-resident layers post no CPU work ($C(0) = 0$), whereas layers with $r_\ell \ge 1$ take $C(r_\ell) = \alpha + \beta r_\ell$. Calibrated inside the serving engine on DeepSeek-V4-Flash, $\alpha = 0.140$\,ms, $\beta = 0.184$\,ms per cold route, and $G_\ell = 0.135$--$0.139$\,ms (\cref{app:derivation}). Because each layer joins both branches at a barrier before the next layer begins (\cref{eq:pass}), its expert branches take $\max\{C(r_\ell), G_\ell\}$, and we call $\sum_\ell \Ex[\max\{C(r_\ell), G_\ell\}]$ the \emph{modeled expert-branch latency} of a decode step.

\textbf{Residency sets and expected cold routes.} Given a static residency set $H_\ell$ in layer $\ell$ and per-step route rates $\rho_{\ell e}$ on a development trace, the expected cold routes in layer $\ell$ equal $\mu_\ell = \sum_{e \notin H_\ell} \rho_{\ell e}$. Let $q_\ell$ denote the probability under the joint per-step routing distribution that layer $\ell$ has at least one cold route, $r_\ell > 0$. We seek a residency set $H = \{H_\ell\}$ that minimizes this modeled expert-branch latency.

\textbf{Rule 1 (Observation~3): keep the experts that remove the most cold routes.} Conditioning on $r_\ell$ yields, whenever $\beta \ge 0$ and $G_\ell \le \alpha + \beta$,
\begin{equation}
  \Ex\bigl[\max\{C(r_\ell), G_\ell\}\bigr] = G_\ell + \beta\mu_\ell + (\alpha - G_\ell)\,q_\ell ,
  \label{eq:overlap}
\end{equation}
for any joint correlation among the $k$ routes of a step (\cref{prop:overlap} covers nonzero $C(0)$ as well). On DeepSeek-V4-Flash, the dispatch residual $\alpha - G_\ell$ is 0.0013--0.0056\,ms, at most 3.04\% of $\beta$, so minimizing expected cold routes $\sum_\ell \mu_\ell$ dominates. When layers on a GPU share equal-size slots and uniform $\beta$, filling each GPU's pool with its highest-$\rho_{\ell e}$ experts minimizes $\sum_\ell \mu_\ell$. Its modeled expert-branch latency lies within $\sum_\ell (\alpha - G_\ell)\,q_\ell$ of the optimal static residency set, bounded by 0.142\,ms per step on a 4,864-slot development pool (\cref{prop:overlap}). When $\beta_\ell$ varies across equal-size layers, ranking experts by $\beta_\ell\,\rho_{\ell e}$ preserves this guarantee (\cref{app:derivation}).

\textbf{Capacity.} Because adding slots never increases cold routes, each pool claims all GPU memory remaining after dense weights, maximum forward-pass workspaces, and the longest development request's KV cache. Growing the pool saves $\beta \sum_\ell \Delta\mu_\ell$ per decode step to leading order, with the $q_\ell$ residual adding at most 3.04\% (\cref{eq:capacity}; 0.90\% on development routes under \cref{eq:capacity} when expanding the pool from 4,864 to 5,702 slots).

\textbf{Rule 2 (Observations~2 and~4): compute a decode step's cold routes on the CPU.} Copying a cold expert on demand exposes 0.535\,ms on an idle PCIe link, whereas running cold routes on the CPU alongside the GPU branch exposes roughly $\beta$ per cold route. Background uploads pay off only when future hits amortize transfer latency without displacing static slots; \sys therefore keeps the static residency set fixed and limits dynamic uploads to optional borrowed decode slots (\cref{sec:design-decode}).

\textbf{Rule 3 (Observations~1 and~2): switch prefill paths once by pass length.} An $n$-token prefill pass with $r(n)$ cold routes and $u(n)$ distinct cold experts either runs \emph{split execution}, computing resident routes on the GPUs and paying one host-memory weight read per distinct cold expert plus per-route CPU compute across $r(n)$, or takes a pure \emph{upload path}, staging and copying its $u(n)$ distinct cold experts once for GPU execution. Under stationary token routing, expected distinct cold experts $\bar u(n)$ are nondecreasing and discretely concave. Whenever the expected cost difference between these two fixed prefill paths takes the linear-minus-concave form $\phi_0 + \phi_1 n - \phi_2\,\bar u(n)$ with $\phi_0 < 0 < \phi_1, \phi_2$, split execution is cheaper below a unique $n^*$ and upload above it (\cref{prop:switch}). Under this cost form, a single threshold $N_p$ calibrated on development sessions suffices.

\subsection{Split Execution}
\label{sec:design-prefill}

\textbf{Short prefills via split execution.} Forward passes of $2 \le n < N_p$ tokens use split execution, while $n = 1$ runs the decode graph. While each GPU computes resident routes in place from its pool, passes of five or more tokens follow a three-step CPU pipeline: early prefetch of host input activations and routing metadata overlapping GPU resident compute, contiguous packing of cold token rows into a scratch arena for parallel CPU evaluation, and deterministic rank-ordered reduction returned through a pinned host buffer (\cref{sec:eval-short}). On DeepSeek-V4-Flash, this leaves the CPU under one of the six routes per token (\cref{sec:eval-short}) and transfers zero expert weights over PCIe.

\textbf{Long prefills via deduplicated upload.} For $n \ge N_p$ tokens, resident routes still execute from the GPU pool via zero slots without re-upload. On GLM-5.3-Flash and Qwen3.8-Flash-Next, CPU threads gather every touched cold expert per layer into pinned staging buffers and upload each once. On DeepSeek-V4-Flash, each partially resident layer sorts touched cold experts by route count, uploads the highest-hit prefix of $k_u$ experts once, computes the rest on the CPU, and merges results in rank order, picking $k_u$ to minimize the slower branch (15.0\,GB transferred on average); on four-GPU, 48-core passes of 512 to 2,048 tokens, a development-calibrated selector sets $k_u$ inside the upload path without changing $N_p = 512$ (\cref{app:ablation}).

\textbf{All-CPU cold branch on short passes.} Short passes compute every cold route on the CPU rather than uploading selected cold experts per layer. Both options read each cold expert once from host memory, at 87\,GB/s on the CPU versus 25\,GB/s over PCIe, so uploading saves only that expert's per-token CPU compute while contending with the CPU branch for host memory bandwidth and adding per-layer planning and branch merging. Extending the long-prefill per-expert selection to short passes on a 12-request DeepSeek-V4-Flash SWE development slice increased total wait across the seven split-path tool turns from 7.60 to 8.69\,s.

\textbf{Threshold.} We calibrate $N_p = 512$ tokens on development sessions, matching both where llama.cpp's CPU and upload paths cross on our workstation and where buun prefills a development session fastest (\cref{sec:background-insights}).

\subsection{A Decode Graph per GPU}
\label{sec:design-decode}

Dispatching cold experts eagerly from the host~\cite{kamahori2025fiddler} inserts dynamic host launches per token, while static device lookup tables~\cite{wang2026seqmoe} capture a graph when all routes execute on the GPU. KTransformers~\cite{chen2025ktransformers} executes routed experts on the CPU (keeping GPU operator shapes constant) and captures decode via \texttt{cudaLaunchHostFunc} callbacks, and its SGLang \texttt{kt-kernel} backend invokes two host callbacks per MoE layer, including layers whose experts all reside on the GPUs. \sys captures multi-GPU hybrid decode into one device-side CUDA graph per GPU while keeping variable-count CPU execution outside host callbacks and preserving unmodified GPU kernels.

\textbf{Fixed shapes through zero slots.} Because the number of resident routes per token varies across steps, each GPU pool reserves $k$ all-zero expert slots (\emph{zero slots}), one per top-$k$ position, and maps every cold selection in the GPU index table to its position's zero slot. Unmodified upstream GPU kernels thus execute $k$ fixed-shape expert operators in every layer, allowing all layers on a GPU to be captured as one CUDA graph. Zero slots execute only in layers that already have a cold route, where the CPU branch takes at least $\alpha + \beta = 0.324$\,ms, exceeding $2\,G_\ell$, so zero-slot compute hides behind the CPU branch.

\textbf{Device-side mailbox synchronization.} \sys places one cache-line mailbox per layer (plus one per GPU) in pinned host memory mapped into every GPU's address space. Before each decode step, the host increments a step counter and launches each GPU's graph once, chaining adjacent GPUs with CUDA events across mapped activation buffers. In each MoE layer, a publish kernel checks the token's routed IDs against the residency table; when a cold route is present, it writes the activations to mapped host memory and issues a system-scope release request while the GPU runs its resident branch. The host thread claims pending layers in order, runs cold experts on the persistent CPU thread pool, and marks completion; a gather kernel acquires completion before the graph adds the CPU output back. All-resident layers post no request and thus incur neither a CPU job nor a GPU wait, and the graph contains no host callbacks.

\textbf{Adaptive graph granularity.} On four GPUs, 20 to 30 of the 43 layers on DeepSeek-V4-Flash have all routes resident on each decode step, so capturing one graph per GPU eliminates per-segment host launch gaps across those layers (\cref{sec:eval-residency}). On one GPU, where 42.3 of 43 layers incur a cold route per decode step versus 13--23 on four GPUs, \sys falls back to per-segment graphs, cutting server-side decode step time from 58.2 to 50.2\,ms on DeepSeek-V4-Flash and from 32.3 to 24.4\,ms on Qwen3.8-Flash-Next.

\textbf{Borrowed memory.} As an additional optimization in decode, workspace reserved by the capacity test for maximum-length prefills is lent to cold decode experts, exposing up to $b = 136$ borrowed slots per GPU, or 544 slots across four GPUs on DeepSeek-V4-Flash, capped at 1,024 slots per GPU. Each CPU cold hit increments an exponentially decaying score; a background thread initiates at most $a = 3$ chunked uploads per decode step across GPUs, replacing a borrowed expert only when a candidate's score exceeds it by a margin. Uploaded experts become active two decode steps later by updating host and device index-table entries in place at fixed device pointers, preserving captured CUDA graphs, and any incoming prefill reclaims the workspace immediately. GPU upload synchronization adds 0.011--0.012\,ms per step on DeepSeek-V4-Flash while shortening decode by 1.4\,ms on SWE and 0.4\,ms on $\tau^2$-bench (\cref{sec:eval-residency}).

\subsection{Running Diverse MoE Architectures}
\label{sec:design-models}

\begin{table}[t]
  \caption{Mechanisms enabled across evaluated models (\cref{sec:eval-scale,sec:eval-qwen}). Pool: cross-layer GPU pool structure; Short: split short prefill ($N < 512$); Pack: short-pass cold-row packing and early input prefetch; Long: long-prefill CPU prefix split; 1-Graph: whole-GPU decode graph (vs.\ per-segment); Borr.: borrowed decode slots on multi-GPU main runs. \textsuperscript{a}Per-segment decode fallback on one GPU; borrowed memory off on reduced-GPU resource sweeps (one and two GPUs on DeepSeek-V4-Flash, one GPU on Qwen3.8-Flash-Next). \textsuperscript{b}Earlier build of \sys (\cref{sec:eval-qwen}).}
  \label{tab:features}
  \centering
  \footnotesize
  \setlength{\tabcolsep}{2.5pt}
  \begin{tabular}{@{} l l c c c c c @{}}
    \toprule
    Model & Pool & Short & Pack & Long & 1-Graph & Borr. \\
    \midrule
    DeepSeek-V4-Flash & single & \ding{51} & \ding{51} & \ding{51} & \ding{51}\textsuperscript{a} & \ding{51}\textsuperscript{a} \\
    GLM-5.3-Flash & size-grp & \ding{51} & -- & -- & \ding{51} & \ding{51} \\
    Qwen3.8-Flash-Next & size-grp & \ding{51} & \ding{51} & -- & \ding{51} & \ding{51}\textsuperscript{a} \\
    Qwen3-235B-A22B\textsuperscript{b} & size-grp & \ding{51} & -- & -- & -- & -- \\
    DeepSeek-V4.1-Flash & size-grp & -- & -- & -- & -- & -- \\
    \bottomrule
  \end{tabular}
\end{table}

Because residency planning, index tables, split execution, and graph capture operate on per-layer expert tensors rather than model-specific blocks, the same core design generalizes across MoE architectures (\cref{tab:models,tab:features}). On GLM-5.3-Flash~\cite{zai2026glm53flash} and Qwen3.8-Flash-Next~\cite{qwen2026qwen38flashnext}, whose Q4\_K\_M weights mix quantization types across projections (14.2--18.5\,MB per expert on GLM-5.3-Flash), the planner partitions slots by expert byte size and selects the byte-group split that maximizes covered development routes (\cref{app:derivation}). The same planner and cross-layer pool port directly to the 94-layer Qwen3-235B-A22B and the 334.8\,GB DeepSeek-V4.1-Flash, which pairs 191.1\,GB of mixed MXFP4 and 2/3-bit IQ experts with a 135.2\,GB CPU Engram table (\cref{sec:eval-scale}).

\subsection{Implementation and Scope}
\label{sec:impl}

We implement \sys on upstream llama.cpp~\cite{ggml2026llamacpp} with CUDA~12.6, retaining llama.cpp's native CPU and CUDA expert kernels (\cref{app:config} lists all commit hashes and modified line counts). Residency pools and upload staging buffers are pre-allocated at start-up from offline profiles, so no forward pass allocates or resizes them, and background borrowed-memory uploads alternate chunks across both PCIe root complexes.

\textbf{Scope.} \sys targets workstation and single-node deployments where routed experts exceed GPU memory and spill to host DRAM; when all weights fit in GPU HBM, standard GPU-only serving suffices. Static cross-layer pools rely on routing stability between offline refreshes (\cref{sec:eval-scale}), and multi-session serving batches pure decode steps of up to $P$ concurrent tokens while dispatching prefills at per-session thresholds (\cref{app:sessions}).

\section{Evaluation}
\label{sec:eval}

Following our setup (\cref{sec:eval-setup}), we evaluate end-to-end session speed (\cref{sec:eval-main}), short-turn prefill costs (\cref{sec:eval-short}), residency granularity, cost boundaries, and decode graphs (\cref{sec:eval-residency}), and generality across workloads, model scales, hardware budgets, and concurrency (\cref{sec:eval-scale}).

\subsection{Experimental Setup}
\label{sec:eval-setup}

\textbf{Models.} We evaluate five open-weight MoE models up to 335\,GB (\cref{tab:models}): DeepSeek-V4-Flash~\cite{deepseekai2026deepseekv4flash}, GLM-5.3-Flash~\cite{zai2026glm53flash}, and Qwen3.8-Flash-Next~\cite{qwen2026qwen38flashnext} in our main matrix, plus Qwen3-235B-A22B~\cite{qwen2025qwen3instruct2507} and DeepSeek-V4.1-Flash~\cite{deepseekai2026deepseekv41flash} (\cref{sec:eval-scale,sec:eval-qwen}). DeepSeek-V4-Flash, GLM-5.3-Flash, and DeepSeek-V4.1-Flash run on four GPUs; Qwen3.8-Flash-Next runs on two. In every setting, routed experts exceed aggregate GPU memory.

\begin{table}[t]
  \caption{Evaluated models. Experts ($k$): MoE layers $\times$ routed experts per layer (active per token); Routed: routed-expert footprint. \textsuperscript{a}Plus a 135.2\,GB CPU Engram table (334.8\,GB total). \textsuperscript{b}Mixed 2- and 3-bit IQ formats.}
  \label{tab:models}
  \centering
  \footnotesize
  \setlength{\tabcolsep}{2.8pt}
  \begin{tabular}{l c r l}
    \toprule
    Model & Experts ($k$) & Routed & Format \\
    \midrule
    DeepSeek-V4-Flash & 43$\times$256 (6) & 147.2\,GB & MXFP4 \\
    GLM-5.3-Flash & 42$\times$288 (8) & 186.5\,GB & Q4\_K\_M \\
    Qwen3.8-Flash-Next & 48$\times$512 (10) & 80.5\,GB & Q4\_K\_M \\
    Qwen3-235B-A22B & 94$\times$128 (8) & 137.3\,GB & Q4\_K\_M \\
    DeepSeek-V4.1-Flash & 40$\times$384 (6) & 191.1\,GB\textsuperscript{a} & MXFP4, IQ\textsuperscript{b} \\
    \bottomrule
  \end{tabular}
\end{table}

\textbf{Workloads.} We replay seven held-out agent workloads across six agent frameworks and benchmarks (\cref{tab:workloads,app:workloads}). Our main matrix replays OpenHands~\cite{wang2025openhands} sessions resolving SWE-rebench issues~\cite{trofimova2025swerebenchopenhandstrajectories,badertdinov2025swerebench} and $\tau^2$-bench conversations~\cite{agentsuite2026tau2bench,barres2026tau2bench} across airline, retail, and telecom domains with recorded tool outputs: DeepSeek-V4-Flash replays 13 SWE sessions and 60 $\tau^2$-bench tasks. GLM-5.3-Flash and Qwen3.8-Flash-Next replay 60 retokenized $\tau^2$-bench tasks plus three SWE sessions at the 25th, 50th, and 75th turn-count percentiles. Five additional held-out workloads totaling 819 requests per model span SWE-agent via \texttt{mini-swe-agent}~\cite{yang2024sweagent}, Terminal-Bench~2.1~\cite{merrill2026terminalbench,buchanan2026benchmarks} data-science debugging (DSH), Claude Code~2.1.34~\cite{anthropic2026claudecode}, and BFCL~v4~\cite{patil2025bfcl} multi-turn Base and Long (\cref{sec:eval-scale}). On the five cross-framework workloads, each DeepSeek-V4-Flash and Qwen3.8-Flash-Next configuration serves them in fixed order on one or two instances, whereas workload-profiled pools and GLM-5.3-Flash start one instance per workload. \sys's residency pools and prefill threshold never use held-out traces.

\begin{table}[t]
  \caption{Main replayed workloads (\cref{app:workloads} lists five additional cross-framework workloads and DeepSeek-V4.1-Flash slices). Outputs: generated tokens; New: median tokens prefilled per turn. \textsuperscript{a}First 30 turns of two development sessions. \textsuperscript{b}GLM-5.3-Flash / Qwen3.8-Flash-Next.}
  \label{tab:workloads}
  \centering
  \footnotesize
  \setlength{\tabcolsep}{3.4pt}
  \begin{tabular}{l r r r r}
    \toprule
    Set & Sessions & Turns & Outputs & New \\
    \midrule
    \multicolumn{5}{l}{\emph{Software engineering (SWE)}} \\
    \quad development & 6 & 400 & 105,879 & 258 \\
    \quad screening\textsuperscript{a} & 2 & 60 & 9,842 & 335 \\
    \quad held-out & 13 & 893 & 215,341 & 248 \\
    \quad held-out\textsuperscript{b} & 3 & 201 & 33,462 / 39,351 & 196 / 223 \\
    \multicolumn{5}{l}{\emph{$\tau^2$-bench}} \\
    \quad development & 24 & 335 & 32,931 & 135 \\
    \quad held-out & 60 & 916 & 83,687 & 118 \\
    \bottomrule
  \end{tabular}
\end{table}

\begin{table*}[t]
  \caption{\sys finishes every held-out workload fastest with the lowest tool-turn wait and decode latency. Default baseline configurations (\cref{app:config}); Time: total request time; Wait: median tool-turn TTFT; TPOT: median decode latency; best in bold. \textsuperscript{a}Second-largest cache size from development stress tests; the largest ran out of GPU memory on a held-out $\tau^2$-bench first turn.}
  \label{tab:main}
  \centering
  \footnotesize
  \setlength{\tabcolsep}{5pt}
  \begin{tabular}{l rrr @{\hspace{14pt}} rrr}
    \toprule
    & \multicolumn{3}{c}{SWE} & \multicolumn{3}{c}{$\tau^2$-bench} \\
    \cmidrule(lr){2-4}\cmidrule(lr){5-7}
    Engine & Time (s) & Wait (s) & TPOT (ms) & Time (s) & Wait (s) & TPOT (ms) \\
    \midrule
    \multicolumn{7}{l}{\emph{DeepSeek-V4-Flash, four GPUs}} \\
    \chip{ktc}KTransformers & 16,950.1 & 2.36 & 62.8 & 7,739.6 & 1.88 & 61.6 \\
    \chip{layerc}llama.cpp (layers) & 12,215.8 & 2.12 & 47.4 & 5,278.2 & 1.81 & 47.1 \\
    \chip{cachec}llama.cpp expert cache & 9,938.4 & 3.36 & 31.4 & 4,513.1 & 2.77 & 29.8 \\
    \chip{ikc}ik\_llama.cpp & 11,207.8 & 2.61 & 39.6 & 4,795.4 & 2.08 & 36.5 \\
    \chip{buunc}buun & 8,967.1 & 3.37 & 26.8 & 4,216.6 & 2.79 & 25.6 \\
    \rowcolor{oursbg}\chip{oursc}\textbf{\sys} & \textbf{6,581.1} & \textbf{0.82} & \textbf{25.9} & \textbf{2,733.1} & \textbf{0.64} & \textbf{25.2} \\
    \midrule
    \multicolumn{7}{l}{\emph{GLM-5.3-Flash, four GPUs; baselines load via \texttt{mmap}}} \\
    \chip{layerc}llama.cpp (layers) & 5,073.4 & 9.73 & 70.6 & 10,384.6 & 9.21 & 64.0 \\
    \chip{cachec}llama.cpp expert cache & 8,182.6 & 15.03 & 157.6 & 17,408.0 & 13.87 & 128.9 \\
    \chip{buunc}buun & 5,246.1 & 15.44 & 71.7 & 12,323.0 & 13.81 & 58.9 \\
    \rowcolor{oursbg}\chip{oursc}\textbf{\sys} & \textbf{2,216.7} & \textbf{2.45} & \textbf{48.7} & \textbf{4,548.3} & \textbf{2.12} & \textbf{43.7} \\
    \midrule
    \multicolumn{7}{l}{\emph{Qwen3.8-Flash-Next, two GPUs}} \\
    \chip{layerc}llama.cpp (layers) & 1,961.5 & 2.75 & 27.2 & 4,013.8 & 2.36 & 27.0 \\
    \chip{cachec}llama.cpp expert cache & 1,884.5 & 3.82 & 27.6 & 3,988.7 & 3.06 & 22.4 \\
    \chip{ikc}ik\_llama.cpp\textsuperscript{a} & 1,886.5 & 1.68 & 39.2 & 3,938.6 & 1.34 & 35.6 \\
    \chip{buunc}buun & 1,730.4 & 3.99 & 22.2 & 3,807.0 & 3.03 & 19.1 \\
    \rowcolor{oursbg}\chip{oursc}\textbf{\sys} & \textbf{858.4} & \textbf{0.67} & \textbf{16.3} & \textbf{1,775.6} & \textbf{0.52} & \textbf{14.7} \\
    \bottomrule
  \end{tabular}
\end{table*}

\begin{figure*}[t]
  \centering
  \includegraphics{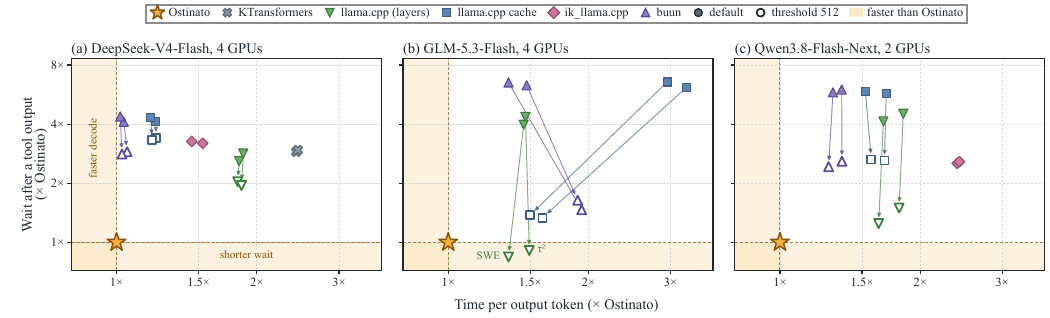}
  \caption{No baseline is faster than \sys in both phases. Every \cref{tab:main} baseline of every held-out cell, placed by its time per output token and its wait after a tool output as multiples of \sys's, which sits at $(1, 1)$; log axes, lower is faster. Each configuration appears once for SWE and once for $\tau^2$-bench. Filled: default configuration; hollow: upload threshold of 512 tokens with matched weight loading; arrows join the two. Shaded: faster than \sys in that phase.}
  \label{fig:tradeoff}
\end{figure*}

\textbf{Hardware.} All experiments run on one workstation with four 24\,GB RTX~4090D GPUs over PCIe~4.0, two Intel Xeon Platinum~8358 CPUs with 64 physical cores, and 512\,GB of DDR4 memory providing 86--88\,GB/s measured host read bandwidth.

\textbf{Baselines.} Every engine passes the same capacity gate requiring the longest development request to complete without out-of-memory failure. We evaluate six CPU--GPU baseline configurations across five engines in three categories (\cref{app:config}). Prior hybrid MoE engines comprise KTransformers~\cite{chen2025ktransformers,kvcacheai2026ktransformers}, run via SGLang's \texttt{kt-kernel} backend on four GPUs with 58 GPU experts per layer under its default uniform placement, which keeps expert IDs 0--57 per layer (\texttt{kt-kernel} also supports optional cross-layer frequency placement), and FreeToken~\cite{yang2026freetoken}, which streams whole layers in prefill and caches slots in decode on one GPU with 32 cores (its DeepSeek-V4-Flash MXFP4 and GGUF expert kernels lack tensor parallelism). Official llama.cpp~\cite{ggml2026llamacpp} provides whole-layer placement and the upstream expert cache~\cite{am17an2026llama}, plus release v1.0.0's auto-tuner. Community MoE forks comprise ik\_llama.cpp~\cite{ikawrakow2026ikllama,samueloliveirads2026add} and buun~\cite{spiritbuun2026cuda}, which warms its cache from profiled routes with decaying-count admission and overlapped decode copies. In \cref{tab:main}, baselines use their default settings and weight loading: GLM-5.3-Flash baselines load via \texttt{mmap} while \sys does not, and on Qwen3.8-Flash-Next all engines use \texttt{mmap} while \sys pins 80.5\,GB of routed experts in host memory. Unpinned controls on 24-request development sweeps still finish 1.46--1.49$\times$ sooner than the fastest aligned baseline (\cref{app:config}). To isolate execution mechanisms, we also evaluate \emph{aligned baselines} at a 512-token upload threshold under matched weight loading, alongside matched-loading ik\_llama.cpp at its default threshold (\cref{sec:eval-short,app:baselines}).

\textbf{Metrics and Numerical Fidelity.} Our primary metric is total request time across a session, with all speedup ratios computed from unrounded measurements. Client-side time to first token after a tool output is a turn's \emph{wait}, with $\tau^2$-bench user turns reported separately, and time per output token (TPOT) divides the interval from first to last token by generated tokens minus one. Every run replays each request with its recorded input and output tokens, and teacher-forced audits on 12 requests per model confirm 97.7--99.3\% top-1 token agreement with official llama.cpp, within 0.5 percentage points of llama.cpp's own placement differences on DeepSeek-V4-Flash and GLM-5.3-Flash and matching cross-commit agreement on Qwen3.8-Flash-Next, with bitwise-identical outputs across repeat runs (\cref{app:correctness}).

\subsection{End-to-End Session Speed}
\label{sec:eval-main}

Across all three main models and both workloads, \sys achieves the lowest total request time and decode latency against every default baseline (\cref{tab:main,fig:teaser}). On DeepSeek-V4-Flash, \sys finishes all 13 SWE sessions and 60 $\tau^2$-bench tasks fastest, 1.36--2.83$\times$ sooner than the five default baselines and 1.54--1.70$\times$ sooner than llama.cpp v1.0.0's auto-tuner (\cref{tab:configs}). Median tool-turn wait drops to 0.82\,s on SWE and 0.64\,s on $\tau^2$-bench (2.6--4.4$\times$ shorter), while p90 wait falls from 3.2--5.5\,s and 2.5--3.9\,s to 1.9 and 1.3\,s (\cref{tab:phase_stitched}). On GLM-5.3-Flash and Qwen3.8-Flash-Next, default speedups reach 2.02--3.83$\times$ in total time and 2.5--6.6$\times$ in tool-turn wait, with the 2.5$\times$ bound set by ik\_llama.cpp on Qwen3.8-Flash-Next (\cref{tab:main}). Even a stitched baseline floor summing the fastest baseline prefill and fastest baseline decode in \cref{tab:main} takes 1.20--2.29$\times$ as long as \sys, including 1.46--1.47$\times$ on Qwen3.8-Flash-Next (\cref{tab:phase_stitched}). Against aligned baselines, \sys stays 1.30--1.94$\times$ faster overall across all three models and 1.17--1.69$\times$ over the aligned stitched floor, with no baseline beating \sys in both phases (\cref{fig:tradeoff,tab:configs}).

Phase totals in \cref{tab:phase_stitched} explain how each baseline family stalls. Against llama.cpp's expert cache and buun on DeepSeek-V4-Flash, \sys cuts total prefill time by 67.5--68.5\% (61.2--61.9\% against ik\_llama.cpp) by running short turns in place and deduplicating long-turn uploads, while also shortening decode by 5.0\% and 2.8\% against buun. Against whole-layer placement, \sys cuts decode time by 45.3\% and 46.9\%, mainly by serving resident routes across all layers instead of evaluating offloaded layers on the CPU, and trims prefill by 50.5\% and 52.4\%. Against KTransformers (\texttt{kt-kernel}), three factors jointly account for the gap: its default uniform placement of expert IDs 0--57 per layer across TP=4 versus 133 frequency-ranked experts on average in \sys, non-AMX AVX-512 CPU prefill kernels on Xeon~8358 hosts, and 256-token KV prefix pages that re-prefill 120--125 extra tokens per turn. \Cref{tab:granularity} quantifies the residency granularity factor at matched GPU bytes.

\subsection{Where Short-Turn Time Goes}
\label{sec:eval-short}

\begin{figure}[t]
  \centering
  \includegraphics{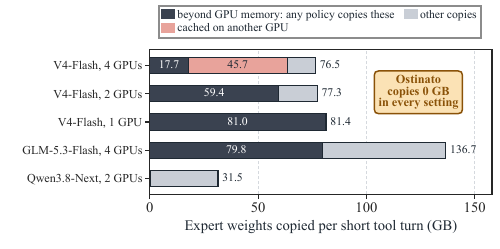}
  \caption{PCIe expert copies per short tool turn (128--511 new tokens) in llama.cpp's cache across models and GPU counts, decomposing cross-GPU re-copies (shown for four-GPU DeepSeek-V4-Flash), misses, and capacity-forced bytes (\cref{tab:bytes}); \sys copies 0\,GB.}
  \label{fig:bytes}
\end{figure}

\begin{figure*}[t]
  \centering
  \includegraphics{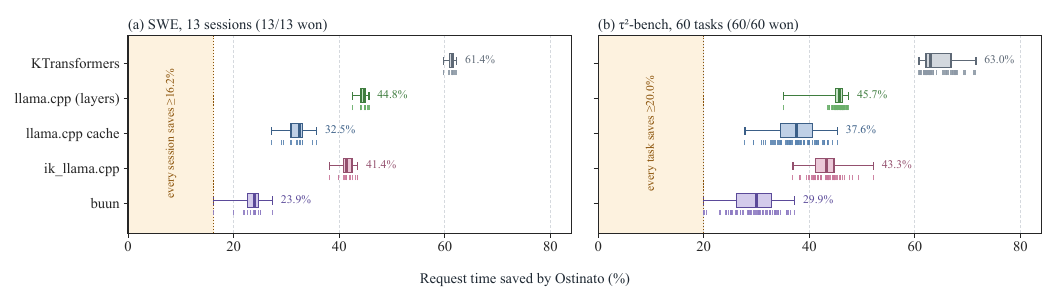}
  \caption{\sys finishes every held-out SWE session~(a) and $\tau^2$-bench task~(b) on DeepSeek-V4-Flash at least 16.2\% sooner than every baseline. llama.cpp's two placements and buun use an upload threshold of 512 tokens, KTransformers and ik\_llama.cpp their defaults. Ticks: individual sessions or tasks; labels: median saving.}
  \label{fig:sessions}
\end{figure*}

\begin{figure}[t]
  \centering
  \includegraphics{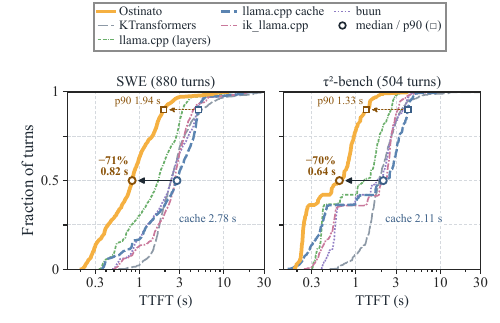}
  \caption{\sys cuts median and tail tool-turn waits on DeepSeek-V4-Flash even when llama.cpp's two placements and buun raise their upload threshold to 512 tokens (KTransformers and ik\_llama.cpp at defaults). Circles and squares mark medians and p90s.}
  \label{fig:ttft}
\end{figure}

\textbf{Byte decomposition.} Byte counters inside llama.cpp expose why default GPU-only caches stall on short tool turns (\cref{fig:bytes,tab:bytes}). Across four-GPU DeepSeek-V4-Flash development turns of 128--511 new tokens, a forward pass touches 93.1\,GB of expert weights on average (154.2\,GB on GLM-5.3-Flash and 46.1\,GB on Qwen3.8-Flash-Next), and llama.cpp's cache copies 76.5\,GB over PCIe: 45.7\,GB re-copy experts cached on another GPU because host-weight dispatches run on GPU~0, 13.0\,GB miss the cache, and 17.7\,GB exceed the GPUs' expert budget (\cref{sec:background-insights}). Capacity-forced copies grow to 59.4\,GB on two GPUs, 81.0\,GB on one GPU, and 79.8\,GB on GLM-5.3-Flash (and 8.6\,GB at Qwen3.8-Flash-Next's configured cache, or 0.5\,GB against all free GPU memory), whereas \sys transfers 0\,GB of expert weights on the split path.

\textbf{Raising upload thresholds to 512 tokens.} Switching baselines to a 512-token upload threshold eliminates short-turn PCIe copies by shifting short prefills entirely onto the CPU (\cref{fig:tradeoff,fig:sessions}). On DeepSeek-V4-Flash, this trims baseline sessions by 2.7--9.3\%, leaving \sys 16.2--45.6\% faster across all 13 SWE sessions and 20.0--47.5\% faster across all 60 $\tau^2$-bench tasks against the three threshold-512 baselines (up to 62.2\% and 71.6\% including KTransformers and ik\_llama.cpp at defaults). Aligned expert caches still wait 2.8--3.4$\times$ longer (2.37--2.78\,s on SWE) because CPUs evaluate all six experts per token while idling resident GPU experts. Aligned layer placement waits 1.9--2.0$\times$ longer and loses all 635 paired short turns. On Qwen3.8-Flash-Next, \sys waits 0.67 and 0.52\,s versus 0.83 and 0.78\,s for aligned layer placement, wins all 296 paired short turns, and finishes sessions 1.69--1.94$\times$ sooner across all threshold-512 baselines (\cref{tab:configs}). Across all 880 SWE and 504 $\tau^2$-bench tool turns on DeepSeek-V4-Flash (\cref{fig:ttft}), \sys cuts median wait by 70.6\% and 69.8\% over aligned llama.cpp's cache and lowers p90 wait from 4.99 and 4.12\,s to 1.94 and 1.33\,s. It also answers user messages in 0.28\,s versus 0.47--1.63\,s (\cref{tab:turntypes}) and sustains 25.5--26.3\,ms/token decode across every SWE context quartile (\cref{fig:turns}).

\textbf{Same-pool forced-upload control.} Fixing \sys's 5,702-expert pool and forcing all prefills onto the deduplicated upload path isolates split execution on the 60-request screening set (\cref{tab:ablation}a). Forcing uploads slows short-turn prefill by 2.10$\times$ on four GPUs, from 33.1 to 69.6\,s (median wait 0.68 to 1.64\,s), and by 1.32$\times$ on one GPU, from 108.9 to 143.5\,s (wait 2.28 to 3.56\,s). Copying cold experts therefore loses on short turns even after static residency deduplicates hits.

\textbf{Marginal per-token wait.} Prefilling 128 to 1,024 new tokens after eight fixed development prefixes (7.4K--39.3K context tokens) against aligned buun (\cref{fig:length}d) isolates token-count scaling. Across the split regime up to 512 new tokens, buun waits 2.54--2.85$\times$ longer (1.63 vs.\ 0.64\,s at 128 tokens): each extra new token adds 2.0\,ms to wait in \sys versus 6.0\,ms in buun because CPU threads evaluate only 0.92--0.95 of the six active experts per token rather than all six. At 1,024 tokens on the upload path, \sys stays 2.39$\times$ faster (2.14 vs.\ 5.12\,s) by keeping resident experts on GPUs and issuing one pinned upload per layer.

\begin{figure*}[t]
  \centering
  \includegraphics{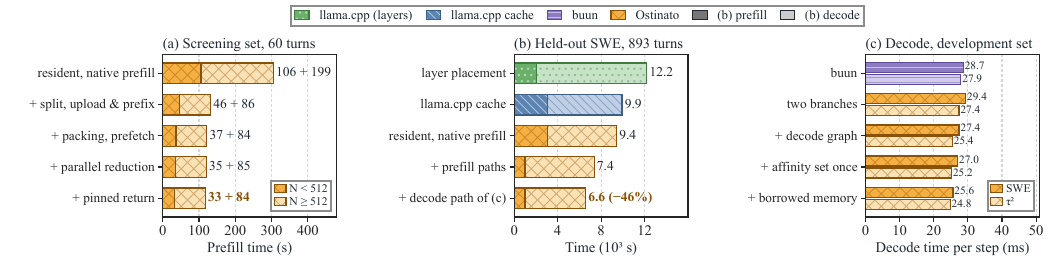}
  \caption{Cumulative stage ablation on DeepSeek-V4-Flash (\cref{tab:ablation}). (a)~Screening-set prefill split below (dark) and from 512 new tokens (light) on the evaluated build. (b)~Held-out SWE prefill and decode totals ($-46\%$ vs.\ layer placement). (c)~Decode step latency on development prefixes across development builds.}
  \label{fig:ablation}
\end{figure*}

\begin{figure*}[t]
  \centering
  \includegraphics{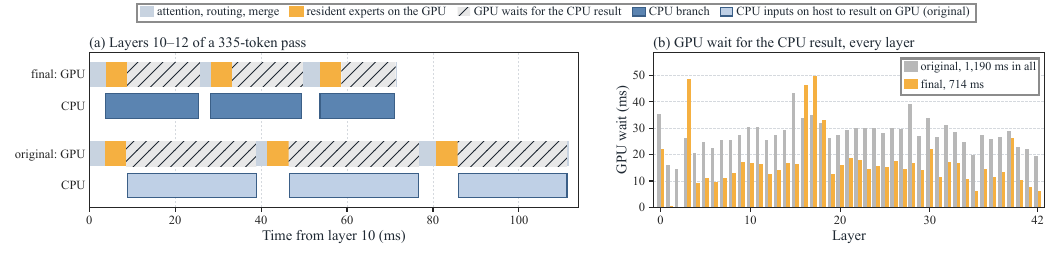}
  \caption{Nsight-joined per-layer timeline of one 335-token DeepSeek-V4-Flash pass on four GPUs before (original) and after (final) short-path overlap optimizations. (a)~Layers 10--12 aligned at layer-10 attention. (b)~GPU wait after resident experts across all 43 layers.}
  \label{fig:timeline}
\end{figure*}

\textbf{Prefill stage ablation and timeline.} In cumulative stage ablations (\cref{fig:ablation}a,b, \cref{tab:ablation}a,b), Stage~1 jointly enables split execution, deduplicated cold uploads, a rule-based long-turn CPU prefix, and long-turn input prefetch, cutting 60-request screening short-turn prefill by 2.29$\times$ (105.9 to 46.2\,s) and long-turn prefill by 2.33$\times$ (199.1 to 85.5\,s). Adding contiguous cold-row packing, early short-turn input prefetch, parallel rank-ordered reduction, and pinned host return cuts short-turn prefill by another 28\% to 33.1\,s (3.20$\times$ cumulative). Nsight timelines of a 335-token pass confirm that early input prefetch overlaps the CPU branch with GPU resident compute across all 43 layers, cutting cumulative GPU wait from 1,190 to 714\,ms and pass latency by 492\,ms (\cref{fig:timeline}). On held-out SWE, the prefill paths cut total prefill time by 67.5\%.

\subsection{When Residency Governs and Decode Graph Enablement}
\label{sec:eval-residency}

\begin{table}[t]
  \caption{Cross-layer expert pools cut cold decode routes by 2.2--8.5$\times$ over whole-layer placement at the same GPU budget. Exact optima over 24 development requests: whole Layers, uniform Model or per-GPU Quota, shared Pool, and Layers/Pool Ratio.}
  \label{tab:granularity}
  \centering
  \footnotesize
  \setlength{\tabcolsep}{2.6pt}
  \begin{tabular}{l l r r r r r}
    \toprule
    & & & \multicolumn{2}{c}{Quota} & & \\
    \cmidrule(lr){4-5}
    Model & & Layers & Model & GPU & Pool & Ratio \\
    \midrule
    DeepSeek-V4-Flash & SWE & 138.0 & 25.3 & 18.6 & \textbf{17.2} & 8.0$\times$ \\
     & $\tau^2$ & 138.0 & 25.3 & 18.3 & \textbf{16.3} & 8.5$\times$ \\
    GLM-5.3-Flash & SWE & 216.0 & 110.2 & 99.2 & \textbf{98.3} & 2.2$\times$ \\
     & $\tau^2$ & 216.0 & 90.2 & 80.3 & \textbf{79.0} & 2.7$\times$ \\
    Qwen3.8-Flash-Next & SWE & 240.0 & 45.5 & 42.7 & \textbf{39.9} & 6.0$\times$ \\
     & $\tau^2$ & 240.0 & 35.8 & 33.7 & \textbf{29.4} & 8.2$\times$ \\
    \bottomrule
  \end{tabular}
\end{table}

\begin{figure*}[t]
  \centering
  \includegraphics{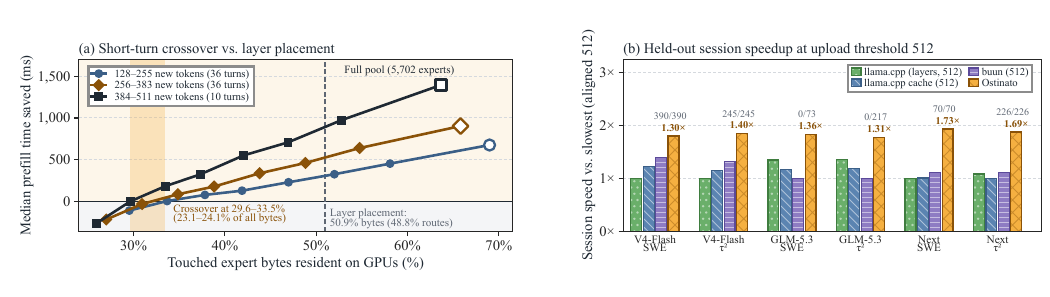}
  \caption{Split execution keeps its short-turn lead over layer placement until the pool shrinks to 23.1--24.1\% of expert bytes. (a)~Median prefill time saved per short turn by split execution over aligned layer placement on DeepSeek-V4-Flash as the pool is pruned; shaded band: the crossover. (b)~Held-out session speed relative to the slowest aligned baseline; labels: \sys's lead over the fastest baseline and paired short-turn wins against layer placement.}
  \label{fig:crossover}
\end{figure*}

\textbf{Residency granularity.} Exact integer placements at matched per-GPU bytes (\cref{tab:granularity}) split the 2.2--8.5$\times$ cold-route reduction of cross-layer pools over whole layers into three steps. First, moving from whole layers to a model-wide uniform expert quota removes 49--85\% of cold decode routes, accounting for 90--97\% of the total reduction by breaking layer confinement. Second, switching to a per-GPU uniform quota removes another 6--28\% of remaining cold routes (26.5--27.9\% on DeepSeek-V4-Flash), because layer-split GPUs hold unequal layer counts (9/12/12/10 on DeepSeek-V4-Flash) alongside dense, embedding, and output weights that cap a global quota at the tightest GPU; this imbalance applies to layer-split rather than tensor-parallel deployments. Third, replacing per-GPU quotas with free cross-layer pools removes a further 6.6--12.8\% of remaining cold routes on skewed DeepSeek-V4-Flash and Qwen3.8-Flash-Next (leaving 7.1--14.6\% excess cold routes for per-GPU quotas and 14.0--55.1\% for model-wide quotas), versus 1.0--1.5\% for per-GPU quotas on flatter-routing GLM-5.3-Flash (0.95\% on SWE and 1.53\% on $\tau^2$-bench; 12.2--14.1\% for model-wide quotas).

\textbf{GLM-5.3-Flash short-turn reversal and cost boundary.} On GLM-5.3-Flash, aligned whole-layer placement waits 0.19--0.39\,s less across tool turns overall (2.06 vs.\ 2.45\,s on SWE; 1.92 vs.\ 2.12\,s on $\tau^2$-bench) and wins all 290 paired short turns of 128--511 tokens (below-512 medians 1.72 vs.\ 2.02\,s and 1.69 vs.\ 1.79\,s). On GLM-5.3-Flash, GPUs hold 38.7\% of routed bytes and flatter routing yields 50.3\% short-turn route coverage (61.4--71.1\% deployed-pool decode coverage on 24 development requests, versus 51.8\% of bytes and 84.6\% short-turn coverage on DeepSeek-V4-Flash); the GLM-5.3-Flash build also runs split execution without short-pass row packing or early input prefetch (\cref{tab:features}). To relate short-turn latency to both byte and route coverage, let a short pass touch $D_\ell$ expert bytes and $R_\ell$ routes in layer $\ell$, with \sys holding byte share $h_\ell$ and route share $p_\ell$ against whole-layer shares $h^L_\ell$ and $p^L_\ell$. Under per-byte host-read cost $\theta_{B\ell}$, per-route CPU compute cost $\theta_{R\ell}$, split setup/return overhead $\sigma$, and branch-overlap difference $\omega$, \cref{app:derivation} derives the short-turn saving
\begin{equation}
  \Delta = \sum_\ell \theta_{B\ell} D_\ell\,(h_\ell - h^L_\ell) + \sum_\ell \theta_{R\ell} R_\ell\,(p_\ell - p^L_\ell) - \sigma + \omega .
  \label{eq:phase}
\end{equation}
Calibrated per model to match each engine's host-read and compute slopes, \cref{eq:phase} reproduces GLM-5.3-Flash's 0/73 and 0/217 short-turn outcomes from its byte and route shares, while a development byte proxy matches Qwen3.8-Flash-Next's 70/70 and 226/226 wins. From 512 tokens upward and on first turns, \sys still answers sooner on GLM-5.3-Flash (5.62 vs.\ 6.73\,s SWE; 5.14 vs.\ 5.82\,s $\tau^2$-bench). Because offloaded layers slow whole-layer decode by 1.35--1.49$\times$, \sys finishes GLM-5.3-Flash sessions 1.31--1.36$\times$ sooner than aligned layer placement and 1.50--1.83$\times$ sooner than aligned caches (\cref{fig:crossover}b).

\textbf{Pool-pruning crossover.} Pruning DeepSeek-V4-Flash's 5,702-expert pool against fixed 21-layer placement across 82 paired short turns directly tests this boundary, locating the crossover at 23.1--24.1\% of expert bytes (95\% bootstrap CI 22.6--24.8\%, \cref{fig:crossover}a). At the crossover, the pruned pool holds fewer touched bytes than layer placement (29.6--33.5\% vs.\ 50.9\%) yet serves more routes (55.7--57.7\% vs.\ 48.8\%; \cref{app:baselines}). Calibrated on DeepSeek-V4-Flash's four larger pruning levels, \cref{eq:phase} matches 7 of 9 held-out bucket-median signs and predicts the 246 turn-level pairs on the three smaller pools at 33.5\,ms RMSE under per-request intercepts (\cref{app:baselines}).

\textbf{Decode graph ablation and remaining CPU cost.} On development prefixes (\cref{fig:ablation}c, \cref{tab:ablation}c; \cref{app:derivation} quantifies CPU--GPU branch overlap), replacing per-segment graphs with one device-side decode graph per GPU saves 1.9\,ms/step, dropping SWE decode from 29.36\,ms (slower than buun's 28.65\,ms) to 27.43\,ms and $\tau^2$-bench from 27.39 to 25.45\,ms. Pinning thread affinity once saves 0.2--0.4\,ms, and borrowed memory saves 1.4\,ms on SWE and 0.4\,ms on $\tau^2$-bench development prefixes; on held-out sessions, borrowed slots absorb 35.7\% of SWE static misses (10.5 of 29.5 misses/step; 5.1 of 20.5 on $\tau^2$-bench). Replacing CPU cold-expert execution with pre-stored outputs on a 120-step SWE prefix lowers median step time from 26.26 to 21.72\,ms, bounding remaining CPU cold-expert computation at 4.54\,ms, or 17\%, per step (\cref{app:bounds}).

\subsection{Generality Across Workloads, Models, Budgets, and Concurrency}
\label{sec:eval-scale}

\begin{figure*}[t]
  \centering
  \includegraphics{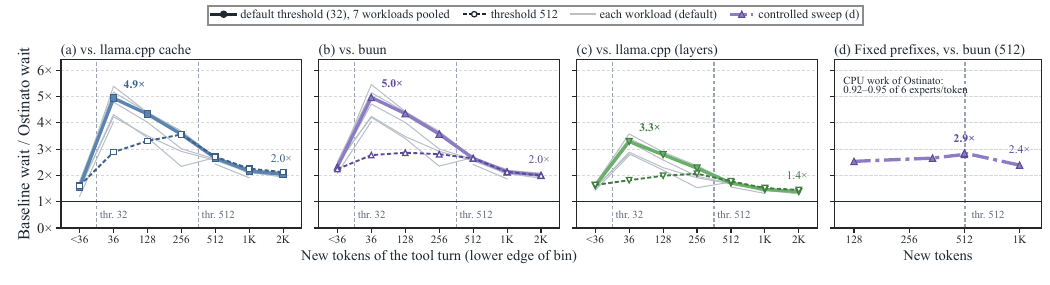}
  \caption{The lead after a tool output tracks the new tokens of a turn in every workload. (a--c)~Median ratio of a baseline's wait to \sys's over 1,968 paired tool turns from seven workloads on DeepSeek-V4-Flash; thin lines: each workload; dashed: the baseline with an upload threshold of 512 tokens (SWE and $\tau^2$-bench only in a and c). (d)~The same ratio against buun at threshold 512 after eight fixed development prefixes.}
  \label{fig:length}
\end{figure*}

\begin{figure}[t]
  \centering
  \includegraphics{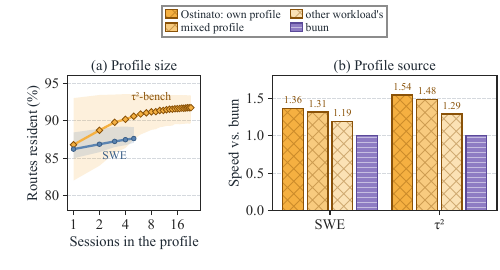}
  \caption{Residency profiles saturate quickly and transfer across workloads on DeepSeek-V4-Flash. (a)~Leave-one-session-out resident decode coverage (p10--p90 band). (b)~Held-out speed relative to buun using matched, mixed, or opposite-workload profiles.}
  \label{fig:profile}
\end{figure}

\begin{figure*}[t]
  \centering
  \includegraphics{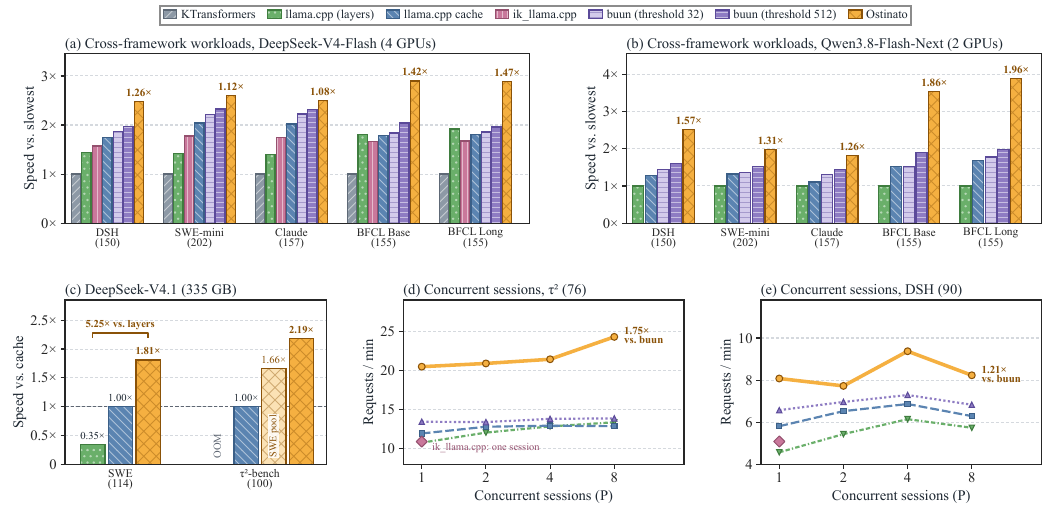}
  \caption{\sys stays fastest on other agent frameworks, on a 335\,GB model, and with concurrent sessions (\cref{tab:newwl,tab:v41,tab:concurrency}). (a,~b)~Five held-out workloads, 819 requests per model, using the transferred SWE pool; labels: speedup over aligned buun at threshold 512. (c)~DeepSeek-V4.1-Flash, speed relative to llama.cpp's cache; layer placement ran out of memory on $\tau^2$-bench. (d,~e)~Fixed-batch throughput with $P$ concurrent sessions; labels: speedup over aligned buun at $P = 8$.}
  \label{fig:scale_conc}
\end{figure*}

\begin{figure*}[t]
  \centering
  \includegraphics{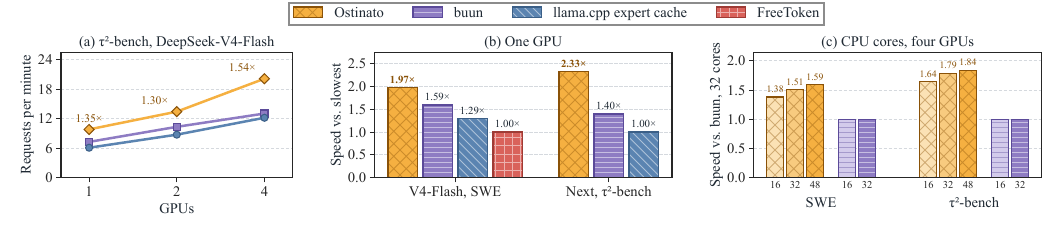}
  \caption{\sys stays fastest across GPU and CPU budgets (\cref{tab:gpus}). (a)~Held-out $\tau^2$-bench throughput on DeepSeek-V4-Flash across 1--4 GPUs (48 cores), labeled against buun. (b)~Single-GPU speed (32 cores) relative to the slowest engine. (c)~Four-GPU development speed across 16--48 cores under identical switches.}
  \label{fig:resources}
\end{figure*}

\textbf{Input-length scaling across seven workloads.} Pooling 1,968 paired tool turns across all seven held-out workloads on DeepSeek-V4-Flash shows that wait speedup tracks new-token length in all seven workloads (\cref{fig:length}a--c). Because llama.cpp evaluates the final four prompt tokens in a separate pass, threshold 32 takes effect at 36--37 new tokens. \sys answers sooner than every default baseline on all 1,479 tool turns of at least 36 tokens (3.6--5.0$\times$ median vs.\ llama.cpp's cache and buun in 36--511 tokens, 2.8--3.6$\times$ vs.\ ik\_llama.cpp, and 1.4--1.7$\times$ vs.\ layer placement from 512 upward); under aligned 512 thresholds, it still cuts median wait in 36--511 tokens by 1.8--2.1$\times$ vs.\ layer placement and 2.8--3.6$\times$ vs.\ llama.cpp's cache and buun (1.4--1.8$\times$ and 2.0--2.7$\times$ from 512 upward, with llama.cpp's two threshold-512 placements evaluated on SWE and $\tau^2$-bench).

\textbf{Residency profile size and transfer.} Small traces saturate fast and transfer across domains (\cref{fig:profile}). Leave-one-session-out decode coverage on $\tau^2$-bench rises from 86.8\% with one session to 91.7\% with 23 (\cref{fig:profile}a). Profiling one $\tau^2$-bench task or SWE session adds only 4.79\% or 1.45\% to held-out total time (staying 1.47$\times$ and 1.34$\times$ faster than buun); an equal-mix profile adds 3.7--4.0\%, and swapping SWE and $\tau^2$-bench profiles adds 14.5--19.7\% while borrowed memory absorbs 49.9--58.3\% of added static misses, keeping \sys 1.19--1.48$\times$ faster than buun (\cref{fig:profile}b).

\textbf{Five cross-framework workloads with a transferred SWE pool.} Replaying 819 held-out requests across Terminal-Bench~2.1 DSH, \texttt{mini-swe-agent}, Claude Code, and BFCL~v4 Base and Long while reusing the static pool profiled on SWE development sessions without re-profiling (the \emph{transferred SWE pool}; \cref{fig:scale_conc}a,b, \cref{tab:newwl}), \sys achieves the lowest total time and median wait across all 30 baseline comparisons on DeepSeek-V4-Flash and all 30 on Qwen3.8-Flash-Next. On DeepSeek-V4-Flash, \sys finishes 1.08--1.47$\times$ sooner than aligned buun, 1.12--2.90$\times$ sooner than default baselines, and 2.48--2.90$\times$ sooner than KTransformers (2.36--6.89$\times$ shorter waits); profiling one BFCL session trims total time by another 7.0\% and 6.6\%. On Qwen3.8-Flash-Next, the transferred SWE pool finishes 1.26--1.96$\times$ sooner than aligned buun (1.29--2.07$\times$ when workload-profiled) and 1.40--3.88$\times$ sooner than default baselines. Across three GLM-5.3-Flash workloads totaling 462 requests (DSH, Claude Code, and BFCL Base), the transferred SWE pool finishes 1.30--4.49$\times$ sooner than every baseline that completes them and waits less than every expert cache (\cref{app:sessions}).

\textbf{Pool portability to 335\,GB and 94 layers.} Porting the cross-layer pool alone to the 334.8\,GB DeepSeek-V4.1-Flash (68\,GB pool with native prefill and per-segment decode; \cref{fig:scale_conc}c, \cref{tab:v41}) finishes 114 SWE requests 1.81$\times$ sooner than llama.cpp's cache (median wait 2.96 vs.\ 6.59\,s) and 5.25$\times$ sooner than 14-layer placement. On 100 $\tau^2$-bench requests, where 14-layer placement runs out of memory on request~89, \sys finishes 2.19$\times$ sooner than the cache (wait 2.89 vs.\ 6.75\,s), and the transferred SWE pool finishes 1.66$\times$ sooner. On the 94-layer Qwen3-235B-A22B (\cref{tab:features,sec:eval-qwen}), \sys cuts $\tau^2$-bench total time by 16.2\% and tool-turn wait by 39.4\% against aligned buun, with 3.1\% slower decode.

\textbf{Fewer GPUs and CPU cores.} \sys retains its lead as GPU and CPU budgets tighten (\cref{fig:resources,tab:gpus}). Across four, two, and one GPU on DeepSeek-V4-Flash $\tau^2$-bench, \sys serves 1.54$\times$, 1.30$\times$, and 1.35$\times$ as many requests per minute as buun, and 1.65$\times$, 1.53$\times$, and 1.61$\times$ as many as llama.cpp's cache (\cref{fig:resources}a). On one GPU with 32 cores, \sys finishes DeepSeek-V4-Flash SWE 1.97$\times$, 1.53$\times$, and 1.24$\times$ faster than FreeToken, llama.cpp's cache, and buun, and Qwen3.8-Flash-Next $\tau^2$-bench 2.33$\times$ and 1.66$\times$ faster than the cache and buun (\cref{fig:resources}b). Cutting four-GPU CPU cores under identical switches mainly affects short-turn wait (\cref{fig:resources}c): 32 vs.\ 48 cores adds 15.5--18.4\% to median wait and 3.0--5.6\% to total time, while 16 cores add 58.7--63.6\% to wait, 4--5\% to decode, and 11.9--15.0\% to total time, still finishing 1.39--1.65$\times$ sooner than 16-core buun.

\textbf{Up to eight concurrent sessions.} Measuring fixed-batch completion time across $P \in \{1, 2, 4, 8\}$ concurrent sessions within a 196,608-token context on DeepSeek-V4-Flash (\cref{fig:scale_conc}d,e, \cref{tab:concurrency}), pure decode steps batch at most $P \le 8$ rows per cold expert (below the 11-token break-even), keeping batched decode on the split path. From 222.6\,s ($P = 1$) to 187.7\,s ($P = 8$), \sys completes the 76-request $\tau^2$-bench batch 1.52--1.75$\times$, 1.63--1.89$\times$, and 1.67--1.90$\times$ sooner than aligned buun, llama.cpp's cache, and layer placement, and finishes the 90-request DSH batch with the transferred SWE pool 1.11--1.29$\times$, 1.18--1.39$\times$, and 1.42--1.77$\times$ sooner. At $P = 8$, \sys holds median wait to 1.26\,s on $\tau^2$-bench and 1.32\,s on DSH versus 2.46--7.09\,s across baselines (\cref{app:sessions}).

\FloatBarrier
\section{Related Work}
\label{sec:related}

\textbf{MoE routing and expert compression.} Oracle-MoE trains routers so consecutive tokens activate consistent experts~\cite{zhou2025oraclemoe}, ReMoE fine-tunes only the gate~\cite{zhu2026remoe}, OEA reroutes tokens toward experts already loaded by the batch~\cite{oncescu2026opportunistic}, FloE compresses experts~\cite{zhou2025floe}, and concurrent work Hibana modifies models with shared expert sets and three-bit weights for unified-memory devices~\cite{hibana2026}. \sys leaves model weights and routing unchanged, exploiting native routing skew across agent turns.

\textbf{Offloading experts to host memory.} MoE-Infinity, HOBBIT, and Pre-gated MoE cache or prefetch experts~\cite{xue2024moeinfinity,tang2024hobbit,hwang2024pregated}; recent work separates prefill and decode loading~\cite{yu2026pdscope,yang2026moecore,zhang2025duoservemoe}, partitions GPU memory between experts and KV cache~\cite{zhang2026wisp}, or compresses experts losslessly~\cite{yang2026zipmoe}. Concurrent work Tideway targets consumer-PC agents with a unified host--GPU pool, split prefill/decode layouts, and fixed-slot decode graphs, yet computes every expert on the GPU~\cite{tideway2026}; \sys runs cold routes of short passes on the CPU without copying weights.

\textbf{Computing experts on the CPU.} Beyond Fiddler~\cite{kamahori2025fiddler} and KTransformers~\cite{chen2025ktransformers} (\cref{sec:background-engines,tab:related,app:related}), DALI~\cite{zhu2026dali} chooses per expert between CPU execution and GPU copy, HybriMoE schedules within each layer~\cite{zhong2025hybrimoe}, and concurrent works Apex~\cite{lv2026apex} and Apto~\cite{zhong2026apto} assign experts per pass or from opposite ends of a shared queue. MoE-Lightning~\cite{cao2025moelightning} targets batched server throughput rather than per-turn wait, while Wang et al.~\cite{wang2026achieving} stream layers for long prompts. FreeToken~\cite{yang2026freetoken} pairs an LRU expert cache with streamed prefill layers and hybrid or offload decode, and PowerInfer pins frequently activated dense neurons on the GPU via cross-layer ILP while running cold neurons on the CPU~\cite{song2024powerinfer}. Across three MoE families in llama.cpp, \sys combines static per-GPU cross-layer pools, packed all-CPU cold short prefills, deduplicated long uploads, and zero-slot device-side decode graphs per GPU.

\textbf{Serving LLMs with tool calls.} InferCept manages KV caches of paused tool-calling requests~\cite{abhyankar2024infercept}, AugServe adapts batch budgets to tool states~\cite{wang2026augserve}, Cake overlaps prefix compute with KV loading~\cite{jin2025compute}, prefix and hierarchical caching avoid recomputation~\cite{kwon2023efficient,zheng2024sglang,gim2024prompt,xie2026strata}, and DistServe disaggregates prefill and decode~\cite{zhong2024distserve}. Prefix caching and KV management reduce recomputation or pause stalls across tool calls, while disaggregation and batch scheduling separate prefill and decode interference; \sys cuts the time to execute returning tool-turn tokens on a CPU--GPU workstation.

\section{Conclusion}
\label{sec:conclusion}

We presented \sys, a CPU--GPU MoE engine for local agents that reuses static cross-layer GPU pools in place across every forward pass. Running sparse cold routes on the CPU in decode and short tool turns (with zero expert-weight copies), deduplicating long-prompt uploads, and capturing multi-GPU decode in one device-side zero-slot CUDA graph per GPU cuts median tool-turn waits by 2.5--6.6$\times$ and finishes SWE and $\tau^2$-bench sessions 1.36--3.83$\times$ faster than default baselines (1.30--1.94$\times$ aligned). \sys finishes sessions fastest across seven agent workloads, one to four GPUs, and up to eight concurrent sessions, while its cross-layer pool speeds up 335\,GB DeepSeek-V4.1-Flash by 1.81--2.19$\times$ over llama.cpp's expert cache. More broadly, expert residency and execution must be co-designed: cross-layer pools determine how many routes GPU memory covers, while resident-in-place execution and device-side synchronization determine whether every forward pass realizes that coverage.

\FloatBarrier

\newpage
\appendix
\crefalias{section}{appendix}
\onecolumn
\phantomsection\label{app:start}   %
\setlength{\parfillskip}{0pt plus 0.8\linewidth}  %

The appendices begin with extended empirical results and conclude with analytical bounds and cost-model derivations. \Cref{app:sessions,app:breakdown,app:baselines,sec:eval-qwen,app:ablation} report per-session and per-domain breakdowns, five cross-framework agent workloads, multi-session concurrency up to eight slots, aligned upload thresholds, two additional models up to 335\,GB, and the numbers behind our resource and ablation figures. \Cref{app:correctness,app:workloads,app:config,app:related} audit numerical precision and cached prefixes, characterize the agent workloads, document engine configurations, and map the design space of CPU--GPU MoE engines. \Cref{app:bounds,app:derivation} state the weight-cut bounds of \cref{sec:background-insights}, empirical schedule controls, and cost-model derivations of \cref{sec:design-model}.

\vspace{4pt}
\noindent{\small
\newcommand{\apptocrow}[3]{\cref{#1} & \hyperref[#1]{\textcolor{black}{#2}}\hspace{4pt}\textcolor{black!35}{\dotfill} & \pageref{#1} \\}%
\renewcommand{\arraystretch}{1.12}%
\begin{tabularx}{\textwidth}{@{} l X r @{}}
  \toprule
  \multicolumn{2}{@{}l}{\textbf{Contents of Appendices}} & \textbf{Page} \\
  \midrule
  \apptocrow{app:sessions}{Per-Session Savings, Domains, Cross-Framework Workloads, and Multi-Session Concurrency}{}
  \apptocrow{app:breakdown}{Prefill, Decode, and Waits by Turn Type}{}
  \apptocrow{app:baselines}{Aligned Upload Thresholds and Weight Loading Across Models}{}
  \apptocrow{sec:eval-qwen}{Qwen3-235B-A22B and DeepSeek-V4.1-Flash}{}
  \apptocrow{app:ablation}{Numbers Behind the Byte, Resource, and Ablation Figures}{}
  \apptocrow{app:correctness}{Correctness, Numerical Precision, and Prefix Audit}{}
  \apptocrow{app:workloads}{Workloads}{}
  \apptocrow{app:config}{Experimental Configurations}{}
  \apptocrow{app:related}{Design Space of CPU--GPU MoE Engines}{}
  \apptocrow{app:bounds}{Analytical Bounds, Cost-Model Envelopes, and Schedule Controls}{}
  \apptocrow{app:derivation}{Cost Model Derivations}{}
  \bottomrule
\end{tabularx}}
\vspace{4pt}

\section{Per-Session Savings, Domains, Cross-Framework Workloads, and Multi-Session Concurrency}
\label{app:sessions}

\Cref{fig:sessions} (\cref{sec:eval-short}) shows the request time saved by \sys on every held-out session and task of DeepSeek-V4-Flash against the five baselines plotted there (llama.cpp's two placements and buun at upload threshold 512, KTransformers and ik\_llama.cpp at their defaults): \sys wins all 13 SWE sessions (saving 16.2--45.6\% against threshold-512 baselines and up to 62.2\% overall) and all 60 $\tau^2$-bench tasks (saving 20.0--47.5\%, up to 71.6\% overall). Splitting $\tau^2$-bench by domain against default baselines (\cref{tab:domains}), \sys finishes the airline, retail, and telecom tasks 1.50--1.58$\times$ sooner than buun (the fastest baseline in each domain) with the shortest median tool-turn wait throughout.

\begin{table}[!ht]
  \caption{\sys is the fastest engine in every $\tau^2$-bench domain. Held-out tasks of DeepSeek-V4-Flash on four GPUs, with the llama.cpp-based baselines in their default configuration: total request time (s) and median wait after a tool output (s); best values in bold.}
  \label{tab:domains}
  \centering
  \small
  \setlength{\tabcolsep}{5pt}
  \begin{tabular}{l rr rr rr}
    \toprule
    & \multicolumn{2}{c}{Airline (20 tasks)} & \multicolumn{2}{c}{Retail (20 tasks)} & \multicolumn{2}{c}{Telecom (20 tasks)} \\
    \cmidrule(lr){2-3}\cmidrule(lr){4-5}\cmidrule(lr){6-7}
    Engine & Total & Wait & Total & Wait & Total & Wait \\
    \midrule
    \chip{layerc}llama.cpp (layers) & 1,639.7 & 2.21 & 1,394.7 & 2.25 & 2,243.8 & 0.42 \\
    \chip{cachec}llama.cpp expert cache & 1,403.8 & 3.39 & 1,236.3 & 3.48 & 1,872.9 & 0.42 \\
    \chip{buunc}buun & 1,303.3 & 3.39 & 1,166.7 & 3.51 & 1,746.6 & 0.61 \\
    \rowcolor{oursbg}\chip{oursc}\textbf{\sys} & \textbf{869.5} & \textbf{0.88} & \textbf{738.8} & \textbf{0.94} & \textbf{1,124.9} & \textbf{0.26} \\
    \bottomrule
  \end{tabular}
\end{table}

\textbf{Five cross-framework agent workloads.} \Cref{tab:newwl} details the 819 held-out requests behind \cref{sec:eval-scale}: Terminal-Bench~2.1 data-science debugging (DSH, 150 requests), \texttt{mini-swe-agent} (202), Claude Code (157), and BFCL~v4 Base (155) and Long (155). On four-GPU DeepSeek-V4-Flash, \sys with the transferred SWE pool achieves the lowest total request time and median tool-turn wait in all 30 baseline comparisons, finishing 1.08--1.47$\times$ sooner than aligned buun, 1.12--2.90$\times$ sooner than default baselines, and 2.48--2.90$\times$ sooner than KTransformers (with 2.36--6.89$\times$ shorter waits). Excluding one looping \texttt{mini-swe-agent} session that emits 32,771 tokens in a single turn (193 requests remaining), \sys finishes 1.33$\times$ sooner than aligned buun (787.7 vs.\ 1,051.1\,s, versus up to 1,423.4\,s across default llama.cpp baselines and 2,095.2\,s on KTransformers). Profiling a single BFCL development session raises static decode coverage on DeepSeek-V4-Flash from 79.7\% to 90.3\% on Base and from 78.6\% to 88.7\% on Long, cutting total time by another 7.0\% and 6.6\%. On two-GPU Qwen3.8-Flash-Next, \sys with the transferred SWE pool wins all 30 baseline comparisons in both total time and median wait, finishing 1.26--3.88$\times$ sooner overall (1.57$\times$, 1.31$\times$, 1.26$\times$, 1.86$\times$, and 1.96$\times$ vs.\ aligned buun, and 1.64$\times$, 1.30$\times$, 1.29$\times$, 2.03$\times$, and 2.07$\times$ with workload-profiled pools); per turn across all seven Qwen3.8-Flash-Next workloads, \sys answers sooner than every \cref{tab:main,tab:newwl} default baseline on all 923 tool turns of at least 36 new tokens, while aligned layer placement answers sooner on 45 of the 417 such turns in the five added workloads, by a median of 24\,ms. Across the three workloads evaluated on four-GPU GLM-5.3-Flash (462 requests: DSH, Claude Code, and BFCL Base), \sys with the transferred SWE pool finishes 1.41--3.18$\times$, 1.30--3.28$\times$, and 1.30--4.49$\times$ sooner than the baselines that complete them (1.32--4.98$\times$ overall with workload-profiled pools); per turn, \sys waits less than every expert-cache baseline, whereas aligned layer placement answers sooner on tool turns below 512 new tokens and \sys answers sooner from 512 tokens upward (\cref{sec:eval-residency}).

\begin{table}[!ht]
  \caption{\sys finishes every completed cross-framework workload pair sooner (total request time, s), both with the transferred SWE pool and with a workload-specific pool, and waits less after tool outputs (median, s) in every pair on DeepSeek-V4-Flash and Qwen3.8-Flash-Next and against every expert-cache baseline (llama.cpp's expert cache and buun) on GLM-5.3-Flash; requests in parentheses; dashes: not run; \textsuperscript{a}ran out of GPU memory on request 61; best values in bold.}
  \label{tab:newwl}
  \centering
  \footnotesize
  \setlength{\tabcolsep}{3.6pt}
  \begin{tabular}{l rr rr rr rr rr}
    \toprule
    & \multicolumn{2}{c}{DSH (150)} & \multicolumn{2}{c}{SWE-mini (202)} & \multicolumn{2}{c}{Claude (157)} & \multicolumn{2}{c}{BFCL Base (155)} & \multicolumn{2}{c}{BFCL Long (155)} \\
    \cmidrule(lr){2-3}\cmidrule(lr){4-5}\cmidrule(lr){6-7}\cmidrule(lr){8-9}\cmidrule(lr){10-11}
    Engine & Time & Wait & Time & Wait & Time & Wait & Time & Wait & Time & Wait \\
    \midrule
    \multicolumn{11}{l}{\emph{DeepSeek-V4-Flash, four GPUs, 48 cores}} \\
    \chip{ktc}KTransformers & 2,880.6 & 1.97 & 4,337.2 & 3.16 & 4,525.7 & 2.45 & 1,640.5 & 1.72 & 2,143.4 & 1.68 \\
    \chip{layerc}llama.cpp (layers) & 2,000.4 & 1.83 & 3,039.9 & 1.95 & 3,236.8 & 1.03 & 904.3 & 0.44 & 1,118.3 & 0.46 \\
    \chip{cachec}llama.cpp expert cache & 1,645.3 & 2.74 & 2,126.8 & 2.95 & 2,238.3 & 1.55 & 919.0 & 0.38 & 1,191.0 & 0.39 \\
    \chip{ikc}ik\_llama.cpp & 1,832.8 & 2.03 & 2,442.8 & 2.15 & 2,591.4 & 1.12 & 985.0 & 0.51 & 1,280.9 & 0.51 \\
    \chip{buunc}buun (32) & 1,541.7 & 2.73 & 1,953.5 & 2.95 & 2,034.8 & 1.52 & 893.6 & 0.62 & 1,153.1 & 0.69 \\
    \chip{buunc}buun (512) & 1,463.2 & 1.84 & 1,865.6 & 1.86 & 1,954.5 & 0.83 & 804.6 & 0.62 & 1,094.6 & 0.66 \\
    \rowcolor{oursbg}\chip{oursc}\textbf{\sys} (SWE pool) & 1,159.5 & 0.84 & \textbf{1,671.4} & \textbf{0.69} & 1,809.3 & 0.36 & 566.6 & 0.31 & 744.1 & 0.32 \\
    \rowcolor{oursbg}\chip{oursc}\textbf{\sys} (profiled) & \textbf{1,151.8} & \textbf{0.73} & 1,680.0 & 0.75 & \textbf{1,809.0} & \textbf{0.34} & \textbf{526.9} & \textbf{0.28} & \textbf{694.9} & \textbf{0.29} \\
    \midrule
    \multicolumn{11}{l}{\emph{Qwen3.8-Flash-Next, two GPUs, 48 cores}} \\
    \chip{layerc}llama.cpp (layers, 32) & 1,989.0 & 2.61 & 2,327.0 & 2.57 & 2,105.9 & 1.44 & 1,439.0 & 0.32 & 2,356.5 & 0.33 \\
    \chip{layerc}llama.cpp (layers, 512) & 1,776.8 & 0.82 & 2,054.0 & 0.77 & 1,921.5 & 0.35 & 1,247.8 & 0.31 & 2,148.9 & 0.32 \\
    \chip{cachec}llama.cpp expert cache (32) & 1,560.5 & 3.37 & 1,765.1 & 3.60 & 1,900.3 & 1.85 & 949.8 & 0.52 & 1,403.2 & 0.56 \\
    \chip{cachec}llama.cpp expert cache (512) & 1,398.5 & 1.41 & 1,601.4 & 1.36 & 1,750.7 & 0.63 & 806.0 & 0.49 & 1,242.4 & 0.53 \\
    \chip{buunc}buun (32) & 1,386.6 & 3.20 & 1,708.6 & 3.50 & 1,618.5 & 1.75 & 944.6 & 0.48 & 1,326.6 & 0.50 \\
    \chip{buunc}buun (512) & 1,238.2 & 1.30 & 1,533.0 & 1.24 & 1,457.1 & 0.53 & 757.8 & 0.45 & 1,186.0 & 0.47 \\
    \rowcolor{oursbg}\chip{oursc}\textbf{\sys} (SWE pool) & 788.8 & 0.80 & \textbf{1,170.7} & \textbf{0.61} & 1,158.1 & 0.32 & 407.1 & 0.30 & 606.6 & 0.31 \\
    \rowcolor{oursbg}\chip{oursc}\textbf{\sys} (profiled) & \textbf{755.6} & \textbf{0.69} & 1,179.0 & 0.64 & \textbf{1,133.5} & \textbf{0.32} & \textbf{373.5} & \textbf{0.26} & \textbf{571.9} & \textbf{0.28} \\
    \midrule
    \multicolumn{11}{l}{\emph{GLM-5.3-Flash, four GPUs, 48 cores (DSH, Claude Code, and BFCL Base)}} \\
    \chip{layerc}llama.cpp (layers, mmap, 32) & 5,172.2 & 9.83 & -- & -- & 5,032.3 & \textbf{0.96} & 3,945.0 & 0.85 & -- & -- \\
    \chip{layerc}llama.cpp (layers, no mmap, 512) & 2,848.7 & \textbf{1.90} & -- & -- & 4,022.0 & 0.98 & 1,528.0 & \textbf{0.73} & -- & -- \\
    \chip{cachec}llama.cpp expert cache (mmap, 32) & 6,441.9 & 14.08 & -- & -- & 10,167.1 & 4.97 & \multicolumn{2}{c}{\emph{incomplete}\textsuperscript{a}} & -- & -- \\
    \chip{buunc}buun (mmap, 32) & 4,236.2 & 14.04 & -- & -- & 4,615.3 & 1.52 & 2,628.3 & 1.19 & -- & -- \\
    \chip{buunc}buun (no mmap, 32) & 3,578.5 & 5.55 & -- & -- & -- & -- & -- & -- & -- & -- \\
    \chip{buunc}buun (no mmap, 512) & 3,404.6 & 3.41 & -- & -- & 5,667.3 & 1.91 & 1,146.0 & 1.21 & -- & -- \\
    \rowcolor{oursbg}\chip{oursc}\textbf{\sys} (SWE pool) & 2,023.7 & 2.29 & -- & -- & 3,102.8 & 1.21 & 878.2 & 0.95 & -- & -- \\
    \rowcolor{oursbg}\chip{oursc}\textbf{\sys} (profiled) & \textbf{1,935.0} & 2.29 & -- & -- & \textbf{3,048.0} & 1.12 & \textbf{792.4} & 0.76 & -- & -- \\
    \bottomrule
  \end{tabular}
\end{table}

\textbf{Two concurrent sessions.} Serving two $\tau^2$-bench development sessions concurrently (\cref{tab:twosessions}), teacher-forced \sys finishes both sessions in 57.1\,s versus 82.6\,s for llama.cpp's expert cache (1.45$\times$), cutting median tool-turn wait from 3.75 to 1.32\,s (2.84$\times$); under free generation, \sys finishes in 58.4\,s versus 97.7\,s (1.67$\times$).

\begin{table}[!ht]
  \caption{\sys keeps its lead when it serves two sessions at once. Two $\tau^2$-bench development sessions of DeepSeek-V4-Flash served together on four GPUs (15 requests, 5 after a tool output, 1,775 output tokens): time to finish both sessions (s) and median wait after a tool output (s). Forced: every output token forced to the recorded one; free: free generation with the recorded token budgets. Best values in bold.}
  \label{tab:twosessions}
  \centering
  \small
  \setlength{\tabcolsep}{6pt}
  \begin{tabular}{l rr rr}
    \toprule
    & \multicolumn{2}{c}{Forced} & \multicolumn{2}{c}{Free} \\
    \cmidrule(lr){2-3}\cmidrule(lr){4-5}
    Engine & Finish & Wait & Finish & Wait \\
    \midrule
    \chip{cachec}llama.cpp expert cache & 82.6 & 3.75 & 97.7 & 3.90 \\
    \rowcolor{oursbg}\chip{oursc}\textbf{\sys} & \textbf{57.1} & \textbf{1.32} & \textbf{58.4} & \textbf{1.40} \\
    \bottomrule
  \end{tabular}
\end{table}

\textbf{One to eight concurrent sessions.} \Cref{tab:concurrency} evaluates $P \in \{1, 2, 4, 8\}$ concurrent sessions sharing a 196,608-token KV context (24,576 tokens per slot at $P = 8$) on four-GPU DeepSeek-V4-Flash, reporting fixed-batch completion time. On the 76-request $\tau^2$-bench batch, \sys finishes 1.52--1.75$\times$ sooner than aligned buun, 1.63--1.89$\times$ sooner than llama.cpp's cache, and 1.67--1.90$\times$ sooner than layer placement. On the 90-request DSH batch, \sys with the transferred SWE pool runs standalone and finishes in 668.7, 699.2, 576.2, and 656.3\,s, 1.11--1.29$\times$, 1.18--1.39$\times$, and 1.42--1.77$\times$ sooner than aligned buun, llama.cpp's cache, and layer placement. The baselines share one server startup, and standalone controls of llama.cpp's cache differ from the shared-startup runs by at most 1.15\%. At $P = 8$, \sys holds median tool-turn wait to 1.26\,s on $\tau^2$-bench and 1.32\,s on DSH versus 2.46--7.09\,s. In pure decode, batching $P$ active slots supplies at most $P$ rows per expert per step, trading $P\times$ fewer steps against fixed per-step and per-expert costs. For $P \ge 2$, buun disables host sharing via its documented \texttt{--no-host-share} flag to avoid a shared-mailbox race in its default host-sharing path, and ik\_llama.cpp runs at $P = 1$ only because its sliding-window KV compression supports a single sequence.

\begin{table}[!ht]
  \caption{Multi-session batch completion time (Finish, s) and median tool-turn wait (Wait, s) on DeepSeek-V4-Flash across $P \in \{1, 2, 4, 8\}$ concurrent sessions sharing a 196,608-token context on four GPUs; \textsuperscript{a}host-sharing disabled (\texttt{--no-host-share}) for $P \ge 2$ as documented by \texttt{buun}; dashes: single-sequence or failed multi-slot runs; best values in bold.}
  \label{tab:concurrency}
  \centering
  \footnotesize
  \setlength{\tabcolsep}{4.5pt}
  \begin{tabular}{l rr rr rr rr}
    \toprule
    & \multicolumn{2}{c}{$P = 1$} & \multicolumn{2}{c}{$P = 2$} & \multicolumn{2}{c}{$P = 4$} & \multicolumn{2}{c}{$P = 8$} \\
    \cmidrule(lr){2-3}\cmidrule(lr){4-5}\cmidrule(lr){6-7}\cmidrule(lr){8-9}
    Engine & Finish & Wait & Finish & Wait & Finish & Wait & Finish & Wait \\
    \midrule
    \multicolumn{9}{l}{\emph{$\tau^2$-bench batch (76 requests: 8 first, 29 user, 39 tool)}} \\
    \chip{layerc}llama.cpp (layers) & 423.9 & 2.06 & 378.5 & 2.16 & 355.1 & 2.43 & 341.0 & 2.94 \\
    \chip{cachec}llama.cpp expert cache & 381.9 & 3.13 & 356.3 & 3.23 & 352.6 & 3.60 & 354.3 & 3.89 \\
    \chip{ikc}ik\_llama.cpp & 418.5 & 2.53 & -- & -- & -- & -- & -- & -- \\
    \chip{buunc}buun (32) & 371.6 & 3.17 & -- & -- & -- & -- & -- & -- \\
    \chip{buunc}buun (512)\textsuperscript{a} & 339.3 & 2.34 & 340.4 & 2.76 & 330.2 & 3.01 & 328.7 & 4.17 \\
    \rowcolor{oursbg}\chip{oursc}\textbf{\sys} & \textbf{222.6} & \textbf{0.79} & \textbf{218.2} & \textbf{0.88} & \textbf{212.8} & \textbf{1.10} & \textbf{187.7} & \textbf{1.26} \\
    \midrule
    \multicolumn{9}{l}{\emph{Terminal-Bench DSH batch (90 requests: 8 first, 82 tool; {\upshape\sys} standalone SWE pool)}} \\
    \chip{layerc}llama.cpp (layers) & 1,182.0 & 1.85 & 994.0 & 2.02 & 879.1 & 2.25 & 941.8 & 2.46 \\
    \chip{cachec}llama.cpp expert cache & 929.8 & 2.79 & 827.8 & 2.90 & 786.8 & 3.09 & 858.9 & 3.35 \\
    \chip{ikc}ik\_llama.cpp & 1,062.7 & 2.17 & -- & -- & -- & -- & -- & -- \\
    \chip{buunc}buun (32) & 871.6 & 2.75 & -- & -- & -- & -- & -- & -- \\
    \chip{buunc}buun (512)\textsuperscript{a} & 821.9 & 1.84 & 775.8 & 2.21 & 740.7 & 2.33 & 790.9 & 7.09 \\
    \rowcolor{oursbg}\chip{oursc}\textbf{\sys} & \textbf{668.7} & \textbf{0.86} & \textbf{699.2} & \textbf{0.99} & \textbf{576.2} & \textbf{1.15} & \textbf{656.3} & \textbf{1.32} \\
    \bottomrule
  \end{tabular}
\end{table}

\section{Prefill, Decode, and Waits by Turn Type}
\label{app:breakdown}

\begin{figure}[!ht]
  \centering
  \includegraphics{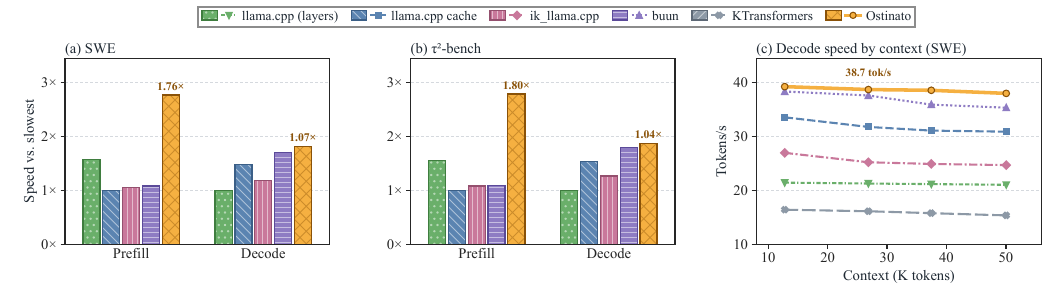}
  \caption{Prefill and decode on DeepSeek-V4-Flash held-out sessions against aligned baselines (\cref{fig:sessions}); higher is better. (a,~b)~Engine-reported prefill and decode speed relative to the slowest engine of each phase (KTransformers omits the phase split), labeled with \sys's speedup over the fastest baseline. (c)~SWE decode throughput by context-length quartile.}
  \label{fig:turns}
\end{figure}

Against aligned baselines on DeepSeek-V4-Flash, \sys leads in both prefill and decode on both main workloads and decodes fastest in every context-length quartile (\cref{fig:turns}). \Cref{tab:phase_stitched} reports prefill and decode totals, p90 and p95 tool-turn waits, paired request win rates, and a non-runnable stitched baseline floor summing the fastest baseline prefill and fastest baseline decode in each cell. Even this stitched floor takes 1.20--2.29$\times$ as long as \sys against default baselines (1.19--2.09$\times$ request-wise, including 1.46--1.47$\times$ aggregate on Qwen3.8-Flash-Next with ik\_llama.cpp setting the fastest baseline prefill) and 1.17--1.69$\times$ against aligned baselines. \sys finishes 99.6--100\% of paired requests sooner than every default baseline in \cref{tab:phase_stitched} (the four exceptions against ik\_llama.cpp on Qwen3.8-Flash-Next $\tau^2$-bench are all first turns); across aligned baselines, the sole exception below 99\% is aligned layer placement on GLM-5.3-Flash $\tau^2$-bench, where \sys still finishes 82\% of paired requests sooner. Across all 1,809 held-out turns of DeepSeek-V4-Flash (\cref{tab:turntypes}), \sys achieves the shortest median wait after tool outputs, user messages, and initial prompts, answering 1.33--9.49$\times$ sooner on first turns exceeding 1,400 tokens.

\begin{table}[!ht]
  \caption{Phase totals, tail waits, per-request win rates and stitched floors of all six held-out main cells. $P$: total prefill time (s); $D$: total decode time (s); p90/p95: 90th and 95th percentile wait for the first token after a tool output (s); Win: share of paired requests where \sys finishes in strictly less wall time than the baseline. \emph{Stitched baseline floor} sums the fastest prefill total among the baselines listed ($\min P$: layer placement or expert cache, ik\_llama.cpp on Qwen3.8-Flash-Next) and the fastest decode total among them ($\min D$, \texttt{buun} except layer placement on GLM-5.3-Flash SWE), and reports the ratio of the aggregate floor ($\min P + \min D$) and the request-wise floor ($\sum_i (\min P_i + \min D_i)$) to \sys's total request time. The aggregate floor takes 1.20--2.29$\times$ as long as \sys and the request-wise floor 1.19--2.09$\times$; with the baselines in their aligned configuration, the aggregate floor takes 1.17--1.69$\times$ as long.}
  \label{tab:phase_stitched}
  \centering
  \footnotesize
  \setlength{\tabcolsep}{4.5pt}
  \begin{tabular}{l r r c r @{\hspace{12pt}} r r c r}
    \toprule
    & \multicolumn{4}{c}{SWE} & \multicolumn{4}{c}{$\tau^2$-bench} \\
    \cmidrule(lr){2-5}\cmidrule(lr){6-9}
    Engine & $P$ (s) & $D$ (s) & Wait p90/p95 (s) & Win & $P$ (s) & $D$ (s) & Wait p90/p95 (s) & Win \\
    \midrule
    \multicolumn{9}{l}{\emph{DeepSeek-V4-Flash, four GPUs (893 SWE / 916 $\tau^2$ requests)}} \\
    \chip{ktc}KTransformers & 3,317 & 13,633 & 5.5 / 7.9 & 100\% & 2,589 & 5,150 & 3.5 / 4.1 & 100\% \\
    \chip{layerc}llama.cpp (layers) & 2,028 & 10,172 & 3.2 / 3.7 & 100\% & 1,381 & 3,893 & 2.5 / 2.6 & 100\% \\
    \chip{cachec}llama.cpp expert cache & 3,095 & 6,830 & 4.8 / 5.5 & 100\% & 2,030 & 2,480 & 3.9 / 4.0 & 99.9\% \\
    \chip{ikc}ik\_llama.cpp & 2,638 & 8,482 & 4.4 / 6.2 & 100\% & 1,697 & 3,046 & 3.1 / 3.4 & 99.9\% \\
    \chip{buunc}buun & 3,104 & 5,852 & 4.8 / 5.4 & 99.9\% & 2,088 & 2,125 & 3.9 / 4.0 & 99.9\% \\
    \rowcolor{oursbg}\chip{oursc}\textbf{\sys} & \textbf{1,005} & \textbf{5,562} & \textbf{1.9 / 2.5} & -- & \textbf{658} & \textbf{2,066} & \textbf{1.3 / 1.5} & -- \\
    \quad \emph{Stitched baseline floor} & 2,028 & 5,852 & \multicolumn{2}{c}{1.20$\times$ / 1.19$\times$} & 1,381 & 2,125 & \multicolumn{2}{c}{1.28$\times$ / 1.26$\times$} \\
    \midrule
    \multicolumn{9}{l}{\emph{GLM-5.3-Flash, four GPUs (201 SWE / 916 $\tau^2$ requests)}} \\
    \chip{layerc}llama.cpp (layers) & 2,834 & 2,237 & 22.2 / 31.7 & 100\% & 6,272 & 4,106 & 11.5 / 19.0 & 100\% \\
    \chip{cachec}llama.cpp expert cache & 2,880 & 5,299 & 18.4 / 19.2 & 100\% & 9,061 & 8,338 & 17.2 / 17.6 & 100\% \\
    \chip{buunc}buun & 2,841 & 2,402 & 18.5 / 19.1 & 100\% & 8,663 & 3,656 & 17.2 / 17.6 & 100\% \\
    \rowcolor{oursbg}\chip{oursc}\textbf{\sys} & \textbf{644} & \textbf{1,567} & \textbf{5.6 / 6.3} & -- & \textbf{1,873} & \textbf{2,628} & \textbf{4.4 / 5.1} & -- \\
    \quad \emph{Stitched baseline floor} & 2,834 & 2,237 & \multicolumn{2}{c}{2.29$\times$ / 1.99$\times$} & 6,272 & 3,656 & \multicolumn{2}{c}{2.18$\times$ / 2.09$\times$} \\
    \midrule
    \multicolumn{9}{l}{\emph{Qwen3.8-Flash-Next, two GPUs (201 SWE / 916 $\tau^2$ requests)}} \\
    \chip{layerc}llama.cpp (layers) & 910 & 1,048 & 6.7 / 10.3 & 100\% & 1,908 & 2,099 & 4.8 / 5.6 & 100\% \\
    \chip{cachec}llama.cpp expert cache & 811 & 1,071 & 5.6 / 6.3 & 100\% & 2,234 & 1,746 & 4.4 / 4.7 & 100\% \\
    \chip{ikc}ik\_llama.cpp & 382 & 1,498 & 2.7 / 3.6 & 100\% & 1,108 & 2,797 & 2.1 / 2.3 & 99.6\% \\
    \chip{buunc}buun & 843 & 873 & 5.9 / 6.5 & 100\% & 2,246 & 1,498 & 4.3 / 4.7 & 100\% \\
    \rowcolor{oursbg}\chip{oursc}\textbf{\sys} & \textbf{223} & \textbf{630} & \textbf{2.0 / 2.6} & -- & \textbf{589} & \textbf{1,142} & \textbf{1.5 / 1.7} & -- \\
    \quad \emph{Stitched baseline floor} & 382 & 873 & \multicolumn{2}{c}{1.46$\times$ / 1.46$\times$} & 1,108 & 1,498 & \multicolumn{2}{c}{1.47$\times$ / 1.46$\times$} \\
    \bottomrule
  \end{tabular}
\end{table}

\begin{table}[!ht]
  \caption{Median time to first token (s) by turn type on DeepSeek-V4-Flash held-out sessions (\cref{fig:ttft}). llama.cpp's two placements and buun use an upload threshold of 512 tokens, KTransformers and ik\_llama.cpp their defaults; turn counts in parentheses; best values in bold.}
  \label{tab:turntypes}
  \centering
  \small
  \setlength{\tabcolsep}{6pt}
  \begin{tabular}{l rrr rr}
    \toprule
    & \multicolumn{3}{c}{After a tool output or user message} & \multicolumn{2}{c}{First turn} \\
    \cmidrule(lr){2-4}\cmidrule(lr){5-6}
    Engine & SWE tool (880) & $\tau^2$ tool (504) & $\tau^2$ user (352) & SWE (13) & $\tau^2$ (60) \\
    \midrule
    \chip{ktc}KTransformers & 2.36 & 1.88 & 1.63 & 32.60 & 7.84 \\
    \chip{layerc}llama.cpp (layers) & 1.66 & 1.24 & 0.47 & 4.57 & 3.82 \\
    \chip{cachec}llama.cpp expert cache & 2.78 & 2.11 & 0.52 & 6.72 & 5.64 \\
    \chip{ikc}ik\_llama.cpp & 2.61 & 2.08 & 0.65 & 10.62 & 6.10 \\
    \chip{buunc}buun & 2.37 & 1.80 & 0.73 & 5.82 & 5.33 \\
    \rowcolor{oursbg}\chip{oursc}\textbf{\sys} & \textbf{0.82} & \textbf{0.64} & \textbf{0.28} & \textbf{3.44} & \textbf{2.68} \\
    \bottomrule
  \end{tabular}
\end{table}

\begin{figure}[!ht]
  \centering
  \includegraphics{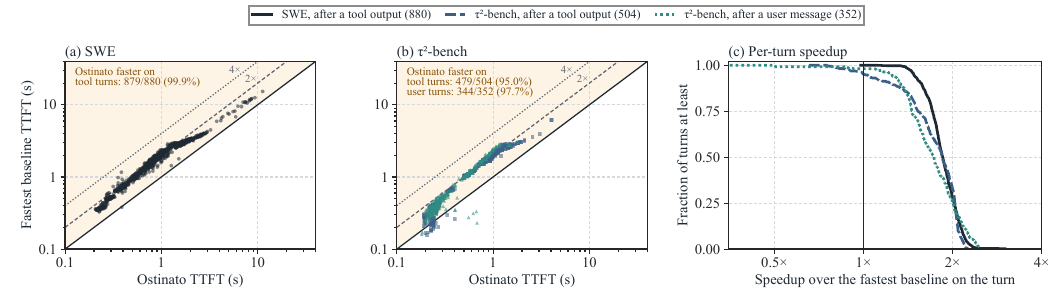}
  \caption{Every post-first held-out turn on DeepSeek-V4-Flash paired against the fastest \cref{tab:turntypes} baseline on that turn. (a,~b)~Time to first token (shaded: \sys faster; dashed and dotted lines: 2$\times$ and 4$\times$). (c)~Fraction of turns achieving at least a given speedup over the per-turn fastest baseline.}
  \label{fig:pairs}
\end{figure}

\textbf{Per-turn comparison against the fastest baseline on each turn.} Pairing each of the 1,736 post-first held-out turns on DeepSeek-V4-Flash against whichever of the five baselines of \cref{tab:turntypes} is fastest on that specific turn (\cref{fig:pairs}), \sys answers sooner on 879 of 880 SWE tool turns (99.9\%), 479 of 504 $\tau^2$-bench tool turns (95.0\%), and 344 of 352 $\tau^2$-bench user turns (97.7\%), with median speedups of 1.85$\times$, 1.85$\times$, and 1.73$\times$ over this per-turn envelope.

\section{Aligned Upload Thresholds and Weight Loading Across Models}
\label{app:baselines}

\Cref{tab:configs} reports held-out aligned runs (512-token upload threshold, matched weight loading) across all three main models, upstream llama.cpp release v1.0.0 with its official auto-tuner fed real development requests and four-GPU capacity on DeepSeek-V4-Flash, held-out ik\_llama.cpp runs under matched weight loading at its own default upload threshold on GLM-5.3-Flash (without \texttt{mmap}) and Qwen3.8-Flash-Next (with default \texttt{mmap}, also in \cref{tab:main}), and 24-request development sweeps on GLM-5.3-Flash without mmap and on Qwen3.8-Flash-Next with mmap (\cref{sec:eval-short,sec:eval-residency}). On the 24-request GLM-5.3-Flash development sweep without mmap, \sys finishes in 232.5\,s on SWE and 152.3\,s on $\tau^2$-bench, 1.48$\times$ and 1.40$\times$ sooner than aligned layer placement; enabling CPU--GPU branch overlap on GLM-5.3-Flash development requests trims median wait by only 16--17\,ms. On held-out sessions against ik\_llama.cpp under matched weight loading at its own default upload threshold, \sys finishes 2.04--2.11$\times$ sooner on GLM-5.3-Flash without \texttt{mmap} (1.70--1.73$\times$ shorter median tool-turn waits) and 2.20--2.22$\times$ sooner on Qwen3.8-Flash-Next under default \texttt{mmap} (2.52--2.57$\times$ shorter waits). Across all six held-out cells (\cref{fig:crossover}b), \sys achieves the lowest total request time (1.30--1.94$\times$ faster than every aligned baseline) and fastest decode, while winning all 635 and all 296 paired short turns (128--511 new tokens) against aligned layer placement on DeepSeek-V4-Flash and Qwen3.8-Flash-Next.

\textbf{Short-turn crossover under pool pruning.} To locate where split execution cedes short turns to fixed whole-layer placement (which keeps 21 of 43 MoE layers on the GPUs using full GPU memory), we prune the DeepSeek-V4-Flash pool across 82 paired tool turns of 128--511 new tokens (\cref{fig:crossover}a). With the full 5,702-expert pool measured in an earlier development run, split execution saves 676--1,394\,ms per turn; at the three smallest pools, \sys wins 74, 38, and 0 of the 82 turns. Interpolating across pruning levels places the crossover at 23.1--24.1\% of total expert bytes across the three length buckets (95\% bootstrap interval 22.6--24.8\% over 20,000 session resamples), where the pool holds 29.6--33.5\% of touched expert bytes and serves 55.7--57.7\% of routes versus 50.9\% of touched bytes and 48.8\% of routes under layer placement. On DeepSeek-V4-Flash, \cref{eq:phase} fitted on the four larger pruning levels matches 7 of 9 held-out bucket-median signs on the three smaller pools, and adding per-request intercepts predicts all 246 turn--level pairs of those 82 turns at 33.5\,ms RMSE. Calibrated per model on GLM-5.3-Flash, \cref{eq:phase} reproduces its 0/73 and 0/217 short-turn outcomes, whereas on Qwen3.8-Flash-Next a development byte proxy aligns with its 70/70 and 226/226 wins.

\begin{table}[!ht]
  \caption{Held-out and 24-request development sweeps under matched weight loading and an upload threshold of 512 tokens, plus upstream \texttt{llama.cpp v1.0.0} on DeepSeek-V4-Flash and ik\_llama.cpp under matched weight loading at its own default upload threshold on GLM-5.3-Flash and Qwen3.8-Flash-Next: total request time (s), median wait after a tool output (s) and time per output token (ms); upload threshold in parentheses (32 is the default of llama.cpp and buun); best values in bold. Dashes: run only with \texttt{mmap} loading. The threshold-512 runs on Qwen3.8-Flash-Next include loading on first use in their first request.}
  \label{tab:configs}
  \centering
  \small
  \setlength{\tabcolsep}{5pt}
  \begin{tabular}{l rrr rrr}
    \toprule
    & \multicolumn{3}{c}{SWE} & \multicolumn{3}{c}{$\tau^2$-bench} \\
    \cmidrule(lr){2-4}\cmidrule(lr){5-7}
    Engine & Total & Wait & TPOT & Total & Wait & TPOT \\
    \midrule
    \multicolumn{7}{l}{\emph{Held-out DeepSeek-V4-Flash: four GPUs, 48 cores, without mmap (893 / 916 requests)}} \\
    \chip{cachec}llama.cpp v1.0.0 (tuner, 32) & 10,165.2 & 3.44 & 31.6 & 4,636.6 & 2.82 & 30.3 \\
    \chip{layerc}llama.cpp (layers, 512) & 11,880.4 & 1.66 & 47.1 & 5,044.5 & 1.24 & 46.7 \\
    \chip{cachec}llama.cpp expert cache (512) & 9,638.7 & 2.78 & 31.5 & 4,357.2 & 2.11 & 30.0 \\
    \chip{buunc}buun (512) & 8,523.9 & 2.37 & 27.3 & 3,824.8 & 1.80 & 25.9 \\
    \rowcolor{oursbg}\chip{oursc}\textbf{\sys} & \textbf{6,581.1} & \textbf{0.82} & \textbf{25.9} & \textbf{2,733.1} & \textbf{0.64} & \textbf{25.2} \\
    \midrule
    \multicolumn{7}{l}{\emph{Held-out GLM-5.3-Flash: four GPUs, 48 cores, without mmap (201 / 916 requests)}} \\
    \chip{layerc}llama.cpp (layers, 512) & 3,007.1 & \textbf{2.06} & 65.5 & 5,956.0 & \textbf{1.92} & 65.2 \\
    \chip{cachec}llama.cpp expert cache (512) & 3,461.1 & 3.27 & 77.5 & 6,842.9 & 2.92 & 65.6 \\
    \chip{ikc}ik\_llama.cpp (1,152) & 4,684.5 & 4.24 & 111.8 & 9,282.1 & 3.61 & 98.3 \\
    \chip{buunc}buun (512) & 4,063.6 & 4.01 & 92.2 & 8,111.4 & 3.10 & 84.5 \\
    \rowcolor{oursbg}\chip{oursc}\textbf{\sys} & \textbf{2,216.7} & 2.45 & \textbf{48.7} & \textbf{4,548.3} & 2.12 & \textbf{43.7} \\
    \midrule
    \multicolumn{7}{l}{\emph{Held-out Qwen3.8-Flash-Next: two GPUs, 48 cores, with mmap (201 / 916 requests)}} \\
    \chip{layerc}llama.cpp (layers, 512) & 1,662.4 & 0.83 & 26.6 & 3,072.2 & 0.78 & 26.4 \\
    \chip{cachec}llama.cpp expert cache (512) & 1,639.8 & 1.74 & 27.4 & 3,335.5 & 1.38 & 23.0 \\
    \chip{ikc}ik\_llama.cpp (1,639) & 1,886.5 & 1.68 & 39.2 & 3,938.6 & 1.34 & 35.6 \\
    \chip{buunc}buun (512) & 1,485.5 & 1.72 & 22.2 & 2,997.4 & 1.27 & 18.7 \\
    \rowcolor{oursbg}\chip{oursc}\textbf{\sys} & \textbf{858.4} & \textbf{0.67} & \textbf{16.3} & \textbf{1,775.6} & \textbf{0.52} & \textbf{14.7} \\
    \midrule
    \multicolumn{7}{l}{\emph{24 development requests, GLM-5.3-Flash: four GPUs, 48 cores, baselines without mmap}} \\
    \chip{cachec}llama.cpp expert cache (32) & -- & -- & -- & 259.6 & 5.79 & 66.6 \\
    \chip{cachec}llama.cpp expert cache (512) & 348.3 & 3.34 & 77.1 & 241.1 & 3.55 & 66.7 \\
    \chip{buunc}buun (32) & -- & -- & -- & 302.6 & 6.02 & 82.4 \\
    \chip{buunc}buun (512) & 395.9 & 4.01 & 90.6 & 279.3 & 3.62 & 82.4 \\
    \chip{layerc}llama.cpp (layers, 32) & -- & -- & -- & 231.1 & 3.92 & 63.0 \\
    \chip{layerc}llama.cpp (layers, 512) & 344.2 & \textbf{2.18} & 65.7 & 213.3 & \textbf{2.21} & 64.7 \\
    \rowcolor{oursbg}\chip{oursc}\textbf{\sys} & \textbf{232.5} & 2.43 & \textbf{48.7} & \textbf{152.3} & 2.46 & \textbf{40.6} \\
    \midrule
    \multicolumn{7}{l}{\emph{24 development requests, Qwen3.8-Flash-Next: two GPUs, 32 cores, all engines with mmap}} \\
    \chip{cachec}llama.cpp expert cache (32) & 235.4 & 4.29 & 27.9 & 134.7 & 3.63 & 24.1 \\
    \chip{cachec}llama.cpp expert cache (512) & 207.3 & 1.81 & 28.0 & 118.4 & 2.07 & 24.0 \\
    \chip{buunc}buun (32) & 217.4 & 3.90 & 22.5 & 125.4 & 3.18 & 20.1 \\
    \chip{buunc}buun (512) & 187.8 & 1.71 & 21.9 & 108.6 & 1.93 & 20.4 \\
    \rowcolor{oursbg}\chip{oursc}\textbf{\sys} & \textbf{97.0} & \textbf{0.94} & \textbf{16.1} & \textbf{63.9} & \textbf{0.99} & \textbf{14.7} \\
    \bottomrule
  \end{tabular}
\end{table}

\Cref{fig:tradeoff} plots each \cref{tab:main} baseline's time per output token against its tool-turn wait, both as multiples of \sys's, across all six held-out cells. At default settings, llama.cpp's cache and buun take as little as 1.02$\times$ \sys's time per token while waiting 4.1--6.6$\times$ as long, ik\_llama.cpp waits 3.2--3.3$\times$ as long on DeepSeek-V4-Flash and 2.52--2.57$\times$ as long (2.40--2.43$\times$ the time per token) on Qwen3.8-Flash-Next, and layer placement waits 2.6--4.5$\times$ as long while taking 1.45--1.87$\times$ the time per token. Raising the upload threshold to 512 tokens shortens baseline waits without closing their decode gap: only aligned layer placement on GLM-5.3-Flash waits less than \sys, 0.84$\times$ and 0.91$\times$ as long, while taking 1.35$\times$ and 1.49$\times$ the time per token.

\section{Qwen3-235B-A22B and DeepSeek-V4.1-Flash}
\label{sec:eval-qwen}

We evaluate cross-layer residency on two larger architectures: the 94-layer Qwen3-235B-A22B (two expert sizes under Q4\_K\_M) and the 334.8\,GB DeepSeek-V4.1-Flash, whose 191.1\,GB of routed experts (40 MoE layers of 384 experts, six active per token) mix MXFP4 with 2- and 3-bit formats (\texttt{IQ2\_XXS}, \texttt{IQ2\_XS}, \texttt{IQ2\_S}, \texttt{IQ3\_XXS}, \texttt{IQ3\_S}) alongside a 135.2\,GB CPU Engram table.

\textbf{Qwen3-235B-A22B.} Running an earlier build of \sys (using llama.cpp's per-segment decode graphs) on the 60 held-out $\tau^2$-bench tasks (\cref{tab:qwen}), \sys finishes 59 tasks sooner than aligned buun, cutting client request time by 16.2\% (14.9\% at the server). Prefill drops by 41.2\% while decode is 3.1\% slower; median wait drops by 39.4\% after tool outputs and 57.8\% after user messages, and median first-turn server prompt processing time drops by 60.0\%.

\textbf{DeepSeek-V4.1-Flash.} On four GPUs with a 38,912-token context (\cref{tab:v41}), \sys pairs a 68\,GB cross-layer expert pool with native prefill and per-segment decode fallback (\cref{sec:design-models}). Across 114 held-out SWE requests and 100 held-out $\tau^2$-bench requests, \sys finishes 1.81$\times$ and 2.19$\times$ sooner than llama.cpp's expert cache (and 1.66$\times$ sooner on $\tau^2$-bench using the transferred SWE pool), whereas 14-layer placement takes 5.25$\times$ as long on SWE and runs out of GPU memory on request~89 of $\tau^2$-bench. In these runs, llama.cpp's cache uses 64,849\,MiB, and first generated tokens match official llama.cpp on all six audited probes (6/6).

\begin{table}[!ht]
  \caption{Held-out $\tau^2$-bench results on the 94-layer Qwen3-235B-A22B (916 turns across 60 tasks, earlier build of \sys) against aligned \texttt{buun} (upload threshold 512); \textsuperscript{a}median server prompt processing time.}
  \label{tab:qwen}
  \centering
  \footnotesize
  \setlength{\tabcolsep}{6pt}
  \begin{tabular}{l r r r}
    \toprule
    Turn type & Turns & buun $\to$ \sys (s) & Change \\
    \midrule
    after a tool output & 504 & 2.75$\to$1.67 & $-39.4\%$ \\
    after a user message & 352 & 1.28$\to$0.54 & $-57.8\%$ \\
    first turn\textsuperscript{a} & 60 & 1.19$\to$0.48 & $-60.0\%$ \\
    \midrule
    total request time & 916 & & $-16.2\%$ \\
    \bottomrule
  \end{tabular}
\end{table}

\begin{table}[!ht]
  \caption{\sys serves the 334.8\,GB DeepSeek-V4.1-Flash 1.81--2.19$\times$ faster than llama.cpp's expert cache on held-out SWE (114 requests) and $\tau^2$-bench (100 requests), four GPUs, 38,912-token context; \textsuperscript{a}ran out of GPU memory on request 89; best values in bold.}
  \label{tab:v41}
  \centering
  \footnotesize
  \setlength{\tabcolsep}{4pt}
  \begin{tabular}{l rrr rrr}
    \toprule
    & \multicolumn{3}{c}{SWE (114)} & \multicolumn{3}{c}{$\tau^2$-bench (100)} \\
    \cmidrule(lr){2-4}\cmidrule(lr){5-7}
    Engine & Time & Wait & TPOT & Time & Wait & TPOT \\
    \midrule
    \chip{layerc}llama.cpp (layers, 14) & 7,161.8 & 6.79 & 265.1 & \multicolumn{3}{c}{\emph{incomplete}\textsuperscript{a}} \\
    \chip{cachec}llama.cpp expert cache & 2,472.3 & 6.59 & 47.1 & 1,439.7 & 6.75 & 46.5 \\
    \chip{oursc}\sys (SWE pool) & -- & -- & -- & 865.8 & 2.89 & 59.1 \\
    \rowcolor{oursbg}\chip{oursc}\textbf{\sys} & \textbf{1,364.4} & \textbf{2.96} & \textbf{40.8} & \textbf{657.8} & \textbf{2.89} & \textbf{37.2} \\
    \bottomrule
  \end{tabular}
\end{table}

\FloatBarrier
\section{Numbers Behind the Byte, Resource, and Ablation Figures}
\label{app:ablation}

\Cref{tab:bytes} lists the short-turn byte totals behind \cref{fig:bytes}, \cref{tab:gpus} lists the total request times behind \cref{fig:resources}a,b, and \cref{tab:ablation} lists all cumulative stages and upload-only controls behind \cref{fig:ablation}.

\textbf{Per-layer timeline of a 335-token short prefill.} \Cref{fig:timeline} (\cref{sec:eval-short}) joins Nsight kernel and copy records with engine hooks across one 335-token pass on four-GPU DeepSeek-V4-Flash before and after short-path overlap optimizations. Without early host-input copies, CPU inputs arrive after resident GPU experts complete in all 43 layers; prefetching host inputs early overlaps the CPU branch with resident GPU experts across every layer, cutting cumulative GPU wait from 1,190 to 714\,ms and pass latency by 492\,ms (from 1,593 to 1,100\,ms, rounded). This single profiled pass pinpoints where the overlap saving arises rather than estimating a population mean.

\textbf{CPU-prefix length selector on the upload path.} Stage~1 of \cref{tab:ablation}a simultaneously activates split execution, deduplicated cold uploads, a rule-based long-turn CPU prefix, and long-turn input prefetch. The held-out configuration on four GPUs and 48 cores replaces only that rule-based CPU-prefix fraction with a cost-model selector calibrated on development prefixes for passes of 512 to 2,048 tokens (other passes and reduced-resource runs retain the rule); the selector operates strictly inside the upload path without changing $N_p = 512$ (\cref{sec:design-prefill}). No stage of \cref{tab:ablation}a uses the selector.

\begin{table}[t]
  \caption{Short tool turns (128--511 new tokens) force GPU-only caches to copy tens of gigabytes over PCIe, whereas \sys copies none. Development-session means (GB): Touched, Copied by llama.cpp's cache, and Forced beyond GPU capacity. \textsuperscript{a}Physical-memory budget of \cref{sec:background-insights}; 8.6\,GB against the 35,840\,MiB configured cache.}
  \label{tab:bytes}
  \centering
  \footnotesize
  \setlength{\tabcolsep}{3.5pt}
  \begin{tabular}{l r r r r}
    \toprule
    Setting & Touched & Copied & Forced & \sys \\
    \midrule
    \multicolumn{5}{l}{\emph{DeepSeek-V4-Flash}} \\
    \quad four GPUs & 93.1 & 76.5 & 17.7 & \textbf{0} \\
    \quad two GPUs & 93.1 & 77.3 & 59.4 & \textbf{0} \\
    \quad one GPU & 93.1 & 81.4 & 81.0 & \textbf{0} \\
    \multicolumn{5}{l}{\emph{GLM-5.3-Flash}} \\
    \quad four GPUs & 154.2 & 136.7 & 79.8 & \textbf{0} \\
    \multicolumn{5}{l}{\emph{Qwen3.8-Flash-Next}} \\
    \quad two GPUs & 46.1 & 31.5 & 0.5\textsuperscript{a} & \textbf{0} \\
    \bottomrule
  \end{tabular}
\end{table}

\begin{table}[!ht]
  \caption{\sys stays the fastest with fewer GPUs and CPU cores. Total request time (s) of the held-out sessions behind \cref{fig:resources}a and b; best values in bold.}
  \label{tab:gpus}
  \centering
  \small
  \setlength{\tabcolsep}{7pt}
  \begin{tabular}{l c c r r r r r}
    \toprule
    Model and workload & Cores & GPUs & \sys & buun & Cache & FreeToken & Speedup \\
    \midrule
    DeepSeek-V4-Flash, $\tau^2$-bench & 48 & 4 & \textbf{2,733} & 4,217 & 4,513 & -- & 1.54$\times$ \\
     &  & 2 & \textbf{4,096} & 5,316 & 6,283 & -- & 1.30$\times$ \\
     &  & 1 & \textbf{5,599} & 7,550 & 9,016 & -- & 1.35$\times$ \\
    \addlinespace[3pt]
    DeepSeek-V4-Flash, SWE & 32 & 1 & \textbf{13,965} & 17,309 & 21,379 & 27,563 & 1.24$\times$ \\
    \addlinespace[3pt]
    Qwen3.8-Flash-Next, $\tau^2$-bench & 32 & 1 & \textbf{2,901} & 4,826 & 6,746 & -- & 1.66$\times$ \\
    \bottomrule
  \end{tabular}
\end{table}

\begin{table}[!ht]
  \caption{Ablation numbers behind \cref{fig:ablation}; each row adds one stage to the row above. (a)~Screening set (60 requests, 5,702 resident experts), last stage in bold; \textsuperscript{a}the upload-only switch also alters $N=1$ decode routing, so only prefill times and short-turn waits are compared. (b)~Held-out SWE (893 requests) against default baselines. (c)~Decode step time on two development prefixes per workload from development builds.}
  \label{tab:ablation}
  \centering
  \footnotesize
  \setlength{\tabcolsep}{4.5pt}
  \begin{tabular}{l r r r r r}
    \multicolumn{6}{l}{\textbf{(a) Screening set (60 requests)}} \\
    \toprule
    & & \multicolumn{2}{c}{Prefill (s)} & & Short-turn \\
    \cmidrule(lr){3-4}
    Configuration & Total (s) & $N < 512$ & $N \ge 512$ & Decode (s) & Wait (s) \\
    \midrule
    resident experts in decode, native prefill & 594.1 & 105.9 & 199.1 & 288.1 & 2.23 \\
    $+$ split path; deduplicated upload, CPU prefix \& input prefetch & 421.8 & 46.2 & 85.5 & 289.3 & 0.97 \\
    $+$ packed cold rows and early short-turn input prefetch & 411.6 & 36.8 & 84.3 & 289.7 & 0.74 \\
    $+$ parallel rank-ordered reduction & 409.3 & 35.0 & 85.0 & 288.5 & 0.72 \\
    $+$ pinned host return & \textbf{408.1} & \textbf{33.1} & \textbf{84.3} & \textbf{289.7} & \textbf{0.68} \\
    \midrule
    \sys, all prefills on upload path (4 GPUs)\textsuperscript{a} & -- & 69.6 & 121.6 & -- & 1.64 \\
    \sys, split vs.\ upload-only (1 GPU)\textsuperscript{a} & -- & 108.9 / 143.5 & 193.7 / 198.5 & -- & 2.28 / 3.56 \\
    \bottomrule
  \end{tabular}

  \vspace{8pt}
  {\footnotesize\setlength{\tabcolsep}{3.5pt}%
  \begin{tabular}[t]{l r r r}
    \multicolumn{4}{l}{\textbf{(b) Held-out SWE (893 requests)}} \\
    \toprule
    Configuration & Total (s) & Prefill (s) & Decode (s) \\
    \midrule
    layer placement & 12,215.8 & 2,028.0 & 10,172.3 \\
    llama.cpp expert cache & 9,938.4 & 3,095.0 & 6,829.8 \\
    resident experts (5,702), native prefill & 9,431.8 & 3,069.8 & 6,346.9 \\
    $+$ size-dependent prefill paths & 7,373.3 & 997.1 & 6,360.2 \\
    $+$ decode graph \& borrowed memory (\textbf{\sys}) & \textbf{6,581.1} & \textbf{1,004.7} & \textbf{5,561.8} \\
    \bottomrule
  \end{tabular}%
  \hspace{14pt}%
  \begin{tabular}[t]{l r r}
    \multicolumn{3}{l}{\textbf{(c) Decode, ms per step}} \\
    \toprule
    Configuration & SWE & $\tau^2$-bench \\
    \midrule
    two branches, per-segment graphs & 29.36 & 27.39 \\
    $+$ decode graph per GPU & 27.43 & 25.45 \\
    $+$ affinity set once & 27.01 & 25.21 \\
    $+$ borrowed memory (\textbf{\sys}) & \textbf{25.63} & \textbf{24.82} \\
    \midrule
    buun & 28.65 & 27.95 \\
    \bottomrule
  \end{tabular}}
\end{table}

\FloatBarrier
\section{Correctness, Numerical Precision, and Prefix Audit}
\label{app:correctness}

\sys evaluates the unmodified routed experts and routing weights of the model while changing where each route executes. Both branches and the upload path invoke llama.cpp's native expert kernels, zero slots contribute exact zeros, and borrowed slots hold verbatim weight copies; because each branch sums its own experts before adding branch outputs and quantizes activations in its native format (\texttt{q8\_0} on CPU, \texttt{q8\_1} on GPU), layer outputs show small numerical differences from branch summation order and activation quantization relative to single-device execution, just as between official llama.cpp placement modes. Across 12 teacher-forced requests per model against official llama.cpp, \sys matches the top-1 token at 99.35\% of 1,537 positions on DeepSeek-V4-Flash (all ten differences are near-ties with top-two gap below 0.5\,nats), with target log-probability absolute differences of median $1.2 \times 10^{-7}$\,nats, p95 0.10\,nats, and maximum 1.73\,nats, matching the 99.41\% top-1 agreement and 0, 0.08, and 1.24\,nats differences between official llama.cpp's layer placement and expert cache. On GLM-5.3-Flash, \sys matches 98.18\% of 935 positions with median $9.9 \times 10^{-5}$, p95 0.195, and maximum 2.20\,nats across 921 paired top-5 positions, versus 98.61\% and $1.1 \times 10^{-4}$, 0.204, and 2.44\,nats within official llama.cpp. On Qwen3.8-Flash-Next, \sys matches 97.75\% of 1,244 positions against upstream llama.cpp at \sys's base commit (p95 0.27\,nats; commits in \cref{app:config}) and 97.35\% against the baseline commit (median $1.7 \times 10^{-5}$, p95 0.44, maximum 3.08\,nats, vs.\ $1.0 \times 10^{-5}$, 0.21, and 1.43\,nats within official llama.cpp), while the two upstream commits agree at 97.03\%. Repeated runs of \sys are bitwise identical on all three models (with deterministic top-$k$ tie-breaking enabled on Qwen3.8-Flash-Next), and repeated held-out $\tau^2$-bench runs vary in total request time by $+2.05\%$ on DeepSeek-V4-Flash, $-0.01\%$ on GLM-5.3-Flash, and $-0.15\%$ on Qwen3.8-Flash-Next.

\textbf{Trajectory and cached-prefix audit.} Across all six main cells (4,043 requests, including 3,844 post-first continuations), every engine replayed the recorded trajectory to completion across all 46 baseline comparisons (31,919 request pairs). Across the 44 llama.cpp-family comparisons (30,110 pairs), actual prefilled and reused token counts differ in 429 pairs, exclusively on first turns; restricted to the remaining 29,681 identical-count pairs, all 44 comparisons preserve their total-time speedup direction, including 1.73--2.21$\times$ on Qwen3.8-Flash-Next SWE, 1.32--3.69$\times$ on GLM-5.3-Flash SWE, and 1.83$\times$, 1.85$\times$, and 1.68$\times$ on the three affected aligned Qwen3.8-Flash-Next cells (aligned layer placement on $\tau^2$-bench and SWE, and aligned buun on $\tau^2$-bench). In the two KTransformers comparisons (1,809 pairs), 256-token KV page reuse and 256-multiple prefill padding alter prefilled or reused counts on 1,800 pairs, averaging 126--128 fewer reused tokens across post-first turns (a median of 120--125 per tool turn; \cref{sec:eval-main}).

\FloatBarrier
\section{Workloads}
\label{app:workloads}

\Cref{fig:workload} (\cref{sec:background-agents}) plots new input tokens and total context lengths across the 1,809 held-out turns of DeepSeek-V4-Flash. Among post-first turns, 85.0\% prefill fewer than 512 new tokens, entering \sys's split path with zero expert-weight copies, while 61.5\% prefill 32 to 511 tokens, where default llama.cpp caches copy every touched expert missing from the computing GPU. Initial prompts prefill a median of 2,032 tokens on SWE and 1,667 on $\tau^2$-bench, and sessions grow to 73.2K context tokens on SWE and 13.2K on $\tau^2$-bench (median 32.4K and 5.4K across turns).

\Cref{tab:workloads_ext} summarizes the five additional cross-framework held-out workloads behind \cref{sec:eval-scale,app:sessions} (819 requests across 92 sessions per model) and the two held-out slices evaluated on the 334.8\,GB DeepSeek-V4.1-Flash (\cref{sec:eval-qwen}). Across all five cross-framework workloads, median post-first prefills stay between 40 and 220 new tokens while contexts extend to 8.3K--46.9K tokens, placing returning turns squarely in the short-turn regime where split execution avoids expert-weight copies.

\begin{table}[!ht]
  \caption{Additional held-out cross-framework workloads (\cref{tab:newwl}) and DeepSeek-V4.1-Flash slices (\cref{tab:v41}). Post-first: continuations reusing a cached prefix; Outputs: generated tokens; New (post-first): median prefilled tokens across all turns (and post-first turns); Max ctx: longest prefix plus input context. \textsuperscript{a}DeepSeek-V4-Flash / Qwen3.8-Flash-Next. \textsuperscript{b}Also run on GLM-5.3-Flash (DSH / Claude Code / BFCL Base): 27,479 / 56,490 / 3,605 outputs, median 203.5 / 34 / 56 new tokens (175 / 34 / 47 post-first), 42.0K / 22.3K / 7.7K max context.}
  \label{tab:workloads_ext}
  \centering
  \footnotesize
  \setlength{\tabcolsep}{5pt}
  \begin{tabular}{l r r r r r r}
    \toprule
    Workload & Sessions & Turns & Post-first & Outputs & New (post-first) & Max ctx \\
    \midrule
    \multicolumn{7}{l}{\emph{Cross-framework held-out workloads\textsuperscript{a} (819 requests per model)}} \\
    \quad Terminal-Bench~2.1 (DSH)\textsuperscript{b} & 11 & 150 & 139 & 33,655 / 32,484 & 195 / 245 (171 / 220) & 45.4K / 46.1K \\
    \quad SWE-agent (\texttt{mini-swe-agent}) & 17 & 202 & 185 & 56,496 / 56,368 & 220 / 273 (158 / 192) & 42.6K / 46.9K \\
    \quad Claude Code~2.1.34\textsuperscript{b} & 12 & 157 & 145 & 64,548 / 62,594 & 40 / 45 (40 / 45) & 22.8K / 23.6K \\
    \quad BFCL~v4 Base\textsuperscript{b} & 26 & 155 & 129 & 10,104 / 7,164 & 62 / 76 (51 / 62) & 8.3K / 8.5K \\
    \quad BFCL~v4 Long & 26 & 155 & 129 & 10,990 / 8,403 & 88 / 103 (63 / 77) & 31.9K / 43.7K \\
    \midrule
    \multicolumn{7}{l}{\emph{DeepSeek-V4.1-Flash held-out slices (38,912-token context)}} \\
    \quad SWE (OpenHands) & 2 & 114 & 112 & 22,072 & 148.5 (139.5) & 38.2K \\
    \quad $\tau^2$-bench & 8 & 100 & 92 & 9,015 & 57.5 (36.0) & 8.9K \\
    \bottomrule
  \end{tabular}
\end{table}

\FloatBarrier
\section{Experimental Configurations}
\label{app:config}

All engines serve DeepSeek-V4-Flash from identical MXFP4 weights with a 98,304-token context (single slot in main runs; YaRN RoPE scaling factor 16 over the native 65,536-token window). Across our three model-specific builds, \sys modifies 93 files with \texttt{+6,299\,/\,-36} lines for DeepSeek-V4-Flash, 82 files with \texttt{+5,110\,/\,-24} lines for Qwen3.8-Flash-Next, and 73 files with \texttt{+4,473\,/\,-15} lines for GLM-5.3-Flash, sharing 48 identical modified files of \texttt{+1,575\,/\,-7} lines; on GLM-5.3-Flash, split execution runs without short-pass row packing and early input prefetch, the long-prefill CPU branch runs only on DeepSeek-V4-Flash, and one- or two-GPU resource sweeps disable borrowed memory (\cref{tab:features}). All llama.cpp-based engines run 48 threads pinned across 48 physical cores (24 per socket) with interleaved NUMA host memory, batch size 4,096, micro-batch 2,048, and an 8,192\,MiB prompt cache; \cref{tab:config} summarizes engine-specific differences. Memory budgets are given in MiB ($2^{20}$ bytes) as passed to each engine, and weight sizes in GB ($10^9$ bytes). \sys builds on upstream llama.cpp commit \texttt{b3fa6995}; official llama.cpp baselines run at commit \texttt{18ab62a8}, except the threshold-512 DeepSeek-V4-Flash runs of layer placement and the expert cache, which use \texttt{b3fa6995}, and the auto-tuner control, which uses release v1.0.0; buun runs at \texttt{1c905459}. On GLM-5.3-Flash and Qwen3.8-Flash-Next (also at 98,304-token context and 48 threads), llama.cpp's expert cache uses 70,912 and 35,840\,MiB (74.4 and 37.6\,GB; the largest capacities passing development stress tests), buun reserves 3,776\,MiB per GPU on GLM-5.3-Flash and its default 640\,MiB on Qwen3.8-Flash-Next, and ik\_llama.cpp uses 52,480 and 28,928\,MiB caches, the second-largest cache sizes tested on development requests, as the largest sizes ran out of GPU memory on a held-out $\tau^2$-bench first turn (default upload thresholds equivalent to 1,152 and 1,639 tokens), loading GLM-5.3-Flash without \texttt{mmap} (\cref{app:baselines}). On GLM-5.3-Flash, default baselines in \cref{tab:main} load via \texttt{mmap} while \sys loads without \texttt{mmap} (\cref{tab:configs} evaluates all aligned baselines without \texttt{mmap}). In \cref{tab:newwl}, GLM-5.3-Flash accordingly reports three buun configurations: default (\texttt{mmap}, threshold 32), a load-mode control on DSH alone (no \texttt{mmap}, threshold 32), and aligned (no \texttt{mmap}, threshold 512). Across five baseline cells in \cref{tab:main,tab:configs} (llama.cpp layer placement on GLM-5.3-Flash SWE and Qwen3.8-Flash-Next SWE/$\tau^2$-bench, llama.cpp cache on GLM-5.3-Flash SWE, and threshold-512 buun on DeepSeek-V4-Flash $\tau^2$-bench), each metric selects the run more favorable to the baseline across two runs; restricting every row to a single run leaves all headline ranges unchanged. On Qwen3.8-Flash-Next, where baselines stall without \texttt{mmap} and copy experts through \texttt{mmap} at 8.3--9.2\,GB/s (vs.\ 24--27\,GB/s on DeepSeek-V4-Flash), all engines load via \texttt{mmap} and \sys additionally pins its 80.5\,GB of routed experts in host memory; disabling those pinned host copies in a paired two-GPU, 32-core control on the 24-request Next development sweeps increases \sys's total time by 36.8\% on SWE (93.8 to 128.2\,s) and 20.5\% on $\tau^2$-bench (60.6 to 73.1\,s, with 79\% and 64\% of the added time in prefills $\ge 512$ tokens), while still finishing 1.46$\times$ and 1.49$\times$ sooner than the fastest aligned baseline in \cref{tab:configs}. KTransformers exposes an SGLang HTTP interface driven by a lightweight client that timestamps the first and last output tokens. Public-code attributes in \cref{tab:related} correspond to official llama.cpp (\texttt{18ab62a8}), buun (\texttt{1c905459}), Fiddler (\texttt{227715bf}), HybriMoE (\texttt{508a85b9}), KTransformers SGLang integration (\texttt{18c0de72}), ik\_llama.cpp (\texttt{13cffc00}), and FreeToken (\texttt{ee43b2a1}).

\begin{table*}[!ht]
  \caption{Engine-specific settings on DeepSeek-V4-Flash; shared settings are documented in \cref{app:config}.}
  \label{tab:config}
  \centering
  \small
  \setlength{\tabcolsep}{5pt}
  \renewcommand{\arraystretch}{1.12}
  \begin{tabularx}{\textwidth}{l >{\raggedright\arraybackslash}X >{\raggedright\arraybackslash}p{4.6cm} >{\raggedright\arraybackslash}X}
    \toprule
    Engine & Experts on the GPUs & Prefill and upload & Other settings \\
    \midrule
    \multicolumn{4}{l}{\emph{Four GPUs, 48 threads on 48 cores}} \\
    \addlinespace[1pt]
    \rowcolor{oursbg}\chip{oursc}\sys & 5,702 most frequently routed experts in a static pool; 544 borrowed slots in decode & split below 512 tokens, upload from 512 & one decode graph per GPU \\
    \addlinespace[2pt]
    \chip{buunc}buun & auto-sized cache admitting experts by decaying access counts, warmed on development routes & upload from 32 (default) or 512 & CPU--GPU overlap on (\texttt{--no-host-share} for $P \ge 2$) \\
    \addlinespace[2pt]
    \chip{cachec}llama.cpp expert cache & 71,936\,MiB (75.4\,GB) cross-layer cache at threshold 32 and 74,048\,MiB (77.6\,GB) at 512, the largest budgets passing each build's capacity test & upload from 32 (default) or 512 & dense weights placed by automatic fit \\
    \addlinespace[2pt]
    \chip{ikc}ik\_llama.cpp & 69,632\,MiB cache sized on development runs & default threshold ($\approx$1,366 tokens) & SWA KV cache (single sequence) \\
    \addlinespace[2pt]
    \chip{ktc}KTransformers (\texttt{kt-kernel}) & expert IDs 0--57 per layer in default uniform placement (optional frequency placement supported; MXFP4 skips online updates) & GPU-prefill threshold 0: splits every prefill in 2,048-token chunks & 4-GPU tensor parallelism; two thread pools; best of seven development configs \\
    \addlinespace[2pt]
    \chip{layerc}llama.cpp (layers) & whole MoE layers by automatic fit (2,048\,MiB reserve per GPU); no cache & CPU layers uploaded from 32 (default) or 512 & none \\
    \midrule
    \multicolumn{4}{l}{\emph{One GPU, 32 threads on 32 cores (16 per socket)}} \\
    \addlinespace[1pt]
    \rowcolor{oursbg}\chip{oursc}\sys & 861 most frequently routed experts & split below 512 tokens, upload from 512 & per-segment decode graphs \\
    \addlinespace[2pt]
    \chip{ftc}FreeToken & 695 expert slots with device-to-device offload & prefill in 2,048-token chunks & none \\
    \bottomrule
  \end{tabularx}
\end{table*}

\textbf{Client versus server timing.} Total request time is measured at the client. On KTransformers, lightweight-client request totals match KTransformers' internal end-to-end timestamps within 0.18\% on SWE and 0.14\% on $\tau^2$-bench; across all other pairs in the main matrix, client- and server-side reductions agree within 0.5 percentage points except Qwen3.8-Flash-Next $\tau^2$-bench against llama.cpp's cache and layer placement ($\approx$1.0 percentage point), as well as buun on Qwen3-235B-A22B (\cref{sec:eval-qwen}).

\textbf{Single-GPU FreeToken setup.} Both \sys and FreeToken run on one RTX~4090D with 32 physical cores (16 per socket), interleaved NUMA memory, and factor-16 YaRN RoPE scaling; although FreeToken's general server supports tensor parallelism, its MXFP4 and GGUF expert kernels on DeepSeek-V4-Flash execute on a single GPU only. \sys keeps 861 resident experts ($N_p = 512$, per-segment decode graphs), while FreeToken serves the same MXFP4 GGUF file as the other engines with 695 expert slots under offload with device-to-device cache hits, the fastest of its four execution strategies on 116 development requests (4,201\,s vs.\ 4,212\,s without device-to-device hits and 8,736--8,745\,s in its two hybrid modes; on the 60-request screening set, FreeToken's native checkpoint path takes 1,394.1\,s versus 1,397.3\,s on the GGUF path, a 0.2\% difference, as its FTW format affects only load time). Because the two engines define internal prefill/decode boundaries differently, we compare end-to-end request time only.

\textbf{Server start-up time.} From initial log output to listening HTTP port, \sys starts in 44.4--316.7\,s across the three main models as it populates static pools and captures decode graphs: 213.9 and 316.7\,s on DeepSeek-V4-Flash SWE and $\tau^2$-bench (2.90$\times$ and 1.89$\times$ faster than llama.cpp's cache at 598--619\,s, vs.\ layer placement 469--604\,s, buun 554--588\,s, ik\_llama.cpp 40--42\,s, and KTransformers 1,800--2,955\,s), 109.4 and 113.0\,s on GLM-5.3-Flash, and 149.2 and 44.4\,s on Qwen3.8-Flash-Next (where the \cref{tab:main} baselines start in 6--112\,s).

\FloatBarrier
\section{Design Space of CPU--GPU MoE Engines}
\label{app:related}

\Cref{tab:related} compares how the CPU--GPU MoE engines of \cref{sec:related} place experts and serve routes.

\begin{table*}[!ht]
  \caption{How CPU--GPU MoE engines place experts and serve routes, read from their public code at the commits of \cref{app:config} and peer papers.}
  \label{tab:related}
  \centering
  \footnotesize
  \setlength{\tabcolsep}{1.7pt}
  \begin{tabular}{@{} l l l l l l @{}}
    \toprule
    Engine & On the GPUs & Residency & Decode misses & Prefill & Decode step \\
    \midrule
    llama.cpp (layers) & whole layers & static & CPU layers on CPU & copy CPU layers & per-segment graphs \\
    llama.cpp cache & experts & on demand & copy before use & one path by threshold & per-segment graphs \\
    ik\_llama.cpp & experts & on demand & copy before use & one path by threshold & per-segment graphs \\
    buun & experts & decaying counts & CPU, staged copies & one path by threshold & per-segment graphs \\
    KTransformers (SOSP'25) & 0 routed experts & static (shared only) & CPU & CPU (AMX) & 1 graph, host callbacks \\
    KTransformers (kt-kernel) & uniform def.; opt.\ freq. & static on MXFP4 & CPU & split or load layers & TP graph, host callbacks \\
    Fiddler & experts & static popularity & CPU & CPU or copy per expert & eager, per expert \\
    HybriMoE & experts & score-based cache & scheduled per layer & GPU (code); hybrid (paper) & eager, per expert \\
    FreeToken & expert slots & on demand (LRU) & CPU/copy (hybrid); copy & stream whole layers & one graph per pass \\
    \rowcolor{oursbg}\sys & expert pool per GPU & static, borrowed & CPU, alongside GPU & split; copy cold $\le 1\times$ & one graph per GPU \\
    \bottomrule
  \end{tabular}
\end{table*}

\FloatBarrier
\section{Analytical Bounds, Cost-Model Envelopes, and Schedule Controls}
\label{app:bounds}

This appendix separates three distinct classes of results: (i)~policy-independent structural lower bounds (\cref{eq:weightcut}, the multi-pass Belady/MIN cache churn bound, and \cref{prop:lease}), which hold for any policy that executes the model's stored weights under its unmodified routing, given the stated capacity and upload-latency constraints; (ii)~cost-model envelopes (\cref{sec:background-example} and \cref{app:derivation}), evaluated under our workstation's calibrated per-expert latencies; and (iii)~empirical schedule controls (the 4.54\,ms stored-output measurement on one 120-step prefix, the $2.10\times$ same-pool upload control, and the 23.1--24.1\% pool-pruning crossover), which measure specific controlled interventions without extrapolating beyond the controlled configuration. For a forward pass touching the set $T$ of experts from an initial GPU set $S$, $W(A)$ denotes the stored bytes of expert set $A$.

\textbf{Weight cut.} Let GPU $j$ compute the touched experts $T_j$ of its layers and start the pass with at most $V_j$ bytes of experts in its memory (its expert budget), and let $X_h$ denote the expert-weight bytes copied from host memory to GPU memory during the pass. If the GPUs compute every route of the pass, each touched expert absent from all GPUs must arrive from host memory, so for every initial set $S$ of experts in GPU memory,
\begin{equation}
  X_h \;\ge\; W(T \setminus S) \;\ge\; \Bigl[W(T) - \sum_j V_j\Bigr]_+ .
  \label{eq:weightcut}
\end{equation}
Counting inter-GPU copies as well, total expert bytes arriving at the GPUs are at least $\sum_j [W(T_j) - V_j]_+$. Setting $\sum_j V_j$ to the capacity of llama.cpp's expert cache yields the capacity-forced bytes of \cref{sec:background-insights,sec:eval-short} (\cref{fig:bytes,tab:bytes}; 15.7\,GB on four-GPU DeepSeek-V4-Flash under the 77.6\,GB threshold-512 cache budget).

\textbf{Across forward passes.} Carrying cache state across alternating prefill and decode passes compounds these copies because no single initial state prepares both phases. On the reconstructed trace of the first 24 DeepSeek-V4-Flash SWE requests (4,000 passes: 53 prefills and 3,947 decode steps across 43 MoE layers; 1,244,489 expert accesses), offline-optimal Belady/MIN caching with bypass on 5,641 global slots (evicting the expert used farthest in the future, starting from an optimal free initial state) loads 1,154.1\,GB (86,321 loads) when carrying state across passes, 27.4\% above the 905.9\,GB (67,757 loads) under free per-phase state resets.

\textbf{Single-path engines.} Under cold-read per-expert costs, any engine executing each pass entirely on the CPU or entirely on the GPUs takes at least the smaller single-path cost regardless of initial residency. In \cref{fig:toy}a, one GPU holds four of eight touched 13.4\,MB experts, so GPU-only prefill copies at least 53.5\,MB (2.14\,ms at 25\,GB/s), while CPU-only prefill pays $0.154 + 0.046\,m$\,ms per $m$-token expert (1.78\,ms across eight experts and 12 routes). Adding 0.11\,ms for three decode steps gives at least 1.89\,ms for any single-path policy ($\ge 2.25$\,ms GPU-only), versus $0.075 + 0.80 + 0.11 = 0.98$\,ms when splitting without overlap.

\textbf{When splitting pays.} Net of split setup and join overheads, splitting a pass beats both single paths whenever the GPU computes resident routes faster than the CPU would and the CPU computes cold routes faster than copying their experts and running them on the GPU. The second condition reverses once enough tokens share each cold expert, motivating the upload path (\cref{prop:switch}). Extrapolating the three-layer costs behind \cref{fig:toy}b (measured up to 32 tokens) linearly beyond 32 tokens, both conditions hold on all 85 forward passes of 2 to 511 tokens captured from a DeepSeek-V4-Flash development session under llama.cpp's expert cache and buun. Whole-layer placement also splits passes across processors, with its net difference against split execution given by \cref{eq:phase}.

\textbf{Granularity.} Let $z_e \in \{0, 1\}$ indicate whether expert $e$ resides on the GPU holding its layer. With $\mathcal{E}_j$ the experts on GPU $j$, $w_e$ the byte size of expert $e$, and $f_e$ its trace route count, an unconstrained per-GPU \emph{expert pool} solves
\begin{equation}
\begin{aligned}
  \underset{z}{\text{maximize}} \quad & \sum_e f_e z_e && \\
  \text{subject to} \quad & \sum_{e \in \mathcal{E}_j} w_e z_e \le V_j && \text{for every GPU } j \quad \text{(GPU memory)},\\
  & z_e \in \{0, 1\} && \text{for every expert } e,
\end{aligned}
\label{eq:placement}
\end{equation}
maximizing resident routes and minimizing cold routes. Whole-layer placement adjoins $z_e = z_{e'}$ within every layer, and a fixed quota adjoins $\sum_{e \in \ell} z_e = \kappa_\ell$ for each layer $\ell$; because both tighten the feasible set of \cref{eq:placement}, neither achieves fewer cold routes than the cross-layer pool. \Cref{tab:granularity} solves these integer programs exactly at matched per-GPU bytes, ranking equal-size experts by $f_e$ and enumerating size-class splits on GLM-5.3-Flash and Qwen3.8-Flash-Next, and compares them against optimal uniform quotas across the whole model or within each GPU.

\textbf{Borrowed memory.} Suppose each GPU lends at most $b$ slots to cold experts during decode, at most $a$ uploads start per decode step across all GPUs, and an upload started after step $t$ becomes usable at step $t + 2$ (with the decode graph waiting if an upload is late); on four-GPU DeepSeek-V4-Flash, $b = 136$ and $a = 3$. Letting $x_{e,t}$ indicate whether cold expert $e$ is available in borrowed memory at step $t$, $y_{e,t}$ whether its upload starts after step $t$, and $r_{e,t}$ the routes to $e$ at step $t$, consider the linear program
\begin{equation}
\begin{aligned}
  \underset{x,\,y}{\text{minimize}} \quad & \sum_t \sum_e r_{e,t}\,(1 - x_{e,t}) && \\
  \text{subject to} \quad & x_{e,t} \le x_{e,t-1} + y_{e,t-2} && \text{for all } e, t \quad \text{(upload latency)},\\
  & \sum_e y_{e,t} \le a && \text{for all } t \quad \text{(uploads per step)},\\
  & \sum_{e \in \mathcal{E}_j} \bigl(x_{e,t} + y_{e,t-1}\bigr) \le b && \text{for all } j, t \quad \text{(borrowed slots)},\\
  & 0 \le x_{e,t},\, y_{e,t} \le 1 && \text{for all } e, t,
\end{aligned}
\label{eq:lease}
\end{equation}
with $x_{e,t} = y_{e,t} = 0$ for $t \le 0$, as borrowed slots start empty at each request.

\begin{proposition}[Borrowed-memory bound]
\label{prop:lease}
The minimum of \cref{eq:lease} is at most the number of cold routes executed on the CPU under any borrowed-memory policy, including one with full knowledge of future routes.
\end{proposition}

\begin{proof}
Fix any policy, setting $x_{e,t} = 1$ if $e$ is available in borrowed memory at step $t$ and $y_{e,t} = 1$ if its upload starts after step $t$, and $0$ otherwise. Availability at step $t$ requires availability at $t - 1$ or an upload started after $t - 2$, giving the first constraint; the per-step upload cap gives the second; and because an in-flight upload ($y_{e,t-1} = 1$) occupies its destination slot until completion without serving routes in step $t$, available and in-flight experts on GPU $j$ share its $b$ borrowed slots, giving the third. Every policy schedule is therefore a feasible integer point of \cref{eq:lease} whose objective equals the cold routes left to the CPU, which exceeds or equals the relaxed minimum.
\end{proof}

On an earlier captured route stream of the first 24 SWE and $\tau^2$-bench development requests on DeepSeek-V4-Flash (with 5,702 resident experts), the LP minimum of \cref{eq:lease} is 14.52 and 12.62 cold routes per step, and feasible integer schedules leave 15.00 and 13.12. Separately, on the first 24 requests served by our final DeepSeek-V4-Flash build (3,947 SWE and 2,893 $\tau^2$-bench decode steps), \sys leaves 19.02 ($75{,}081 / 3{,}947$) and 16.03 ($46{,}380 / 2{,}893$) cold routes per step to the CPU, while the stored-output control on one 120-step SWE development prefix (\cref{sec:eval-residency}) bounds remaining CPU cold-expert computation at 4.54\,ms (17\%) with the decode graph and GPU branch held fixed.

\FloatBarrier
\section{Cost Model Derivations}
\label{app:derivation}

\Cref{tab:notation} summarizes our cost-model notation for single-token decode steps, a fixed layer-to-GPU assignment, and in-engine calibrated branch timings (\cref{sec:design-model}).

\begin{table}[!ht]
  \caption{Notation of the cost model.}
  \label{tab:notation}
  \centering
  \small
  \begin{tabular}{l l}
    \toprule
    Symbol & Meaning \\
    \midrule
    $E$, $k$ & routed experts of an MoE layer; experts that a token selects \\
    $\mathcal{L}_j$, $\kappa_j$ & MoE layers on GPU $j$; expert slots of GPU $j$ \\
    $H_\ell$, $H$ & resident experts of layer $\ell$; the residency set across all layers \\
    $r_\ell$ & cold routes of layer $\ell$ in a decode step \\
    $G_\ell$ & time of the GPU branch of layer $\ell$ in a decode step \\
    $C(r)$ & time of the CPU branch with $r$ cold routes: $\gamma_0$ if $r = 0$, else $\alpha + \beta r$ \\
    $g_\ell$, $\delta_\ell$, $\delta_{\max}$ & $\max\{\gamma_0, G_\ell\}$; $\alpha - g_\ell$; $\max_\ell \delta_\ell$ \\
    $\rho_{\ell e}$ & routes to expert $e$ of layer $\ell$ per decode step \\
    $\mu_\ell(H)$, $q_\ell(H)$ & expected cold routes of layer $\ell$; probability that the layer has any \\
    $M$, $Q$, $F$ & $\sum_\ell \mu_\ell$; $\sum_\ell \delta_\ell q_\ell$; $\beta M + Q$ \\
    $n$, $N_p$; $w$, $B$ & tokens of a prefill pass, threshold; equal-size expert bytes, PCIe upload bandwidth \\
    $r(n)$, $u(n)$, $\bar u(n)$ & cold routes, distinct cold experts, and expected distinct cold experts of an $n$-token pass \\
    \bottomrule
  \end{tabular}
\end{table}

\textbf{Time of a forward pass.} Taking the start of layer $\ell$'s GPU branch as time zero, let $\tau^{\mathrm{pre}}_\ell$ be pre-branch preparation, $G_\ell$ the GPU-branch duration, $d_\ell$ the CPU-branch start offset, $C_\ell$ the CPU-branch duration, and $\tau^{\mathrm{ret}}_\ell$ and $\tau^{\mathrm{join}}_\ell$ the post-branch return and join. With a per-layer barrier before layer $\ell + 1$, a forward pass takes
\begin{equation}
  T_{\mathrm{pass}} = T_{\mathrm{other}} + \sum_\ell \bigl[\tau^{\mathrm{pre}}_\ell + \max\{G_\ell,\, d_\ell + C_\ell\} + \tau^{\mathrm{ret}}_\ell + \tau^{\mathrm{join}}_\ell\bigr],
  \label{eq:pass}
\end{equation}
where $T_{\mathrm{other}}$ collects non-MoE work. Because the maximum applies per layer, two layers with $(C, G) = (8, 1)$ and $(1, 8)$ cost 16 rather than 9. \Cref{prop:overlap} evaluates one decode expert-branch term in \cref{eq:pass}, absorbing $d_\ell$ into $\gamma_0$ and $\alpha$ ($\gamma_0 = 0, g_\ell = G_\ell$ in \cref{sec:design-model}).

\begin{proposition}[Overlapped layers and frequency-ranked residency]
\label{prop:overlap}
Let layer $\ell$ have fixed GPU-branch time $G_\ell$ and CPU-branch time $C(r) = \gamma_0$ for $r = 0$ and $\alpha + \beta r$ for $r \ge 1$, with $\beta \ge 0$ and $G_\ell \le \alpha + \beta$. For any joint distribution of within-step routes,
\begin{equation*}
  \Ex\bigl[\max\{C(r_\ell), G_\ell\}\bigr] = g_\ell + \beta\mu_\ell + \delta_\ell\,q_\ell .
\end{equation*}
If each GPU $j$ shares equal-size expert slots across $\mathcal{L}_j$ without per-layer limits, all layers share $\beta$, and $\delta_\ell \ge 0$, then keeping the $\kappa_j$ most frequently routed experts on each GPU $j$ (set $H_F$) minimizes expected cold routes, and its modeled expert-branch latency exceeds that of an optimal static residency set by at most $Q(H_F)$.
\end{proposition}

\begin{proof}
If $r_\ell = 0$, the layer takes $\max\{\gamma_0, G_\ell\} = g_\ell$; if $r_\ell \ge 1$, $\alpha + \beta r_\ell \ge \alpha + \beta \ge G_\ell$, so it takes $\alpha + \beta r_\ell$. Taking expectations yields
\begin{align*}
  \Ex\bigl[\max\{C(r_\ell), G_\ell\}\bigr] &= (1 - q_\ell)\,g_\ell + \Ex\bigl[(\alpha + \beta r_\ell)\,\mathbf{1}\{r_\ell > 0\}\bigr]\\
  &= (1 - q_\ell)\,g_\ell + \alpha q_\ell + \beta\mu_\ell ,
\end{align*}
proving the first claim, with within-layer co-activation entering solely through $q_\ell$. Summing across layers, modeled expert-branch latency depends on $H$ only via $F(H) = \beta M(H) + Q(H)$, where
\begin{equation*}
  M(H) = \sum_{\ell, e} \rho_{\ell e} - \sum_\ell \sum_{e \in H_\ell} \rho_{\ell e}.
\end{equation*}
With equal-size slots and no per-layer limits, selecting the largest $\rho_{\ell e}$ on each GPU minimizes $M$, so $M(H_F) \le M(H_*)$ for an optimal $H_*$, and
\begin{align*}
  0 \le F(H_F) - F(H_*) &= \beta\bigl(M(H_F) - M(H_*)\bigr) + Q(H_F) - Q(H_*)\\
  &\le Q(H_F) - Q(H_*) \le Q(H_F),
\end{align*}
using $\beta \ge 0$ and $\delta_\ell \ge 0$.
\end{proof}

On DeepSeek-V4-Flash, $g_\ell = G_\ell$ and $\delta_\ell \in [0.0013, 0.0056]$\,ms (at most 3.04\% of $\beta$), bounding $\sum_\ell \delta_\ell \approx 0.152$\,ms per step for any residency set.

\textbf{Trace estimation.} Over $S_0$ decode steps with route indicators $I_{\ell e s} \in \{0, 1\}$ at step $s$,
\begin{equation*}
\begin{alignedat}{2}
  &\rho_{\ell e} = \frac{1}{S_0} \sum_s I_{\ell e s}, &\qquad &r_{\ell s}(H) = \sum_{e \notin H_\ell} I_{\ell e s},\\
  &\mu_\ell(H) = \frac{1}{S_0} \sum_s r_{\ell s}(H), &\qquad &q_\ell(H) = \frac{1}{S_0} \sum_s \mathbf{1}\{r_{\ell s}(H) > 0\},
\end{alignedat}
\end{equation*}
capturing empirical co-activation directly without independence assumptions. On an earlier 4,864-slot DeepSeek-V4-Flash development pool (before the deployed 5,702-slot pool), a layer incurs a cold route on 58.9\% of decode steps on average, giving $Q(H_F) = \sum_\ell \delta_\ell q_\ell \le \delta_{\max}\sum_\ell q_\ell = 0.0056 \times 43 \times 0.589 = 0.142$\,ms per step.

\textbf{Heterogeneous expert sizes and slopes.} When layers differ in expert byte size (as in Q4\_K\_M Qwen3-235B-A22B), keeping the highest-frequency experts within each size class minimizes cold routes in that class, and enumerating feasible class splits maximizes unweighted route coverage (\cref{sec:design-models}; weighting class splits by $\beta_\ell$ extends \cref{prop:overlap} when slopes vary across classes). When equal-size layers differ in slope $\beta_\ell$, ranking experts by $\beta_\ell\,\rho_{\ell e}$ minimizes $\sum_\ell \beta_\ell\,\mu_\ell$, and the $Q(H_F)$ bound of \cref{prop:overlap} carries over.

\textbf{Capacity scaling.} For nested pools $H \subseteq H'$, let $\Delta\mu_\ell = \mu_\ell(H) - \mu_\ell(H')$, $\Delta q_\ell = q_\ell(H) - q_\ell(H')$, $\Delta F = F(H) - F(H')$, and $\Delta F_0 = \beta \sum_\ell \Delta\mu_\ell$. Integer route counts satisfy $0 \le \mathbf{1}\{r_\ell(H) > 0\} - \mathbf{1}\{r_\ell(H') > 0\} \le r_\ell(H) - r_\ell(H')$ step-wise, so $0 \le \Delta q_\ell \le \Delta\mu_\ell$ and, for $\beta > 0$ and $0 \le \delta_\ell \le \delta_{\max}$,
\begin{equation}
  \Delta F_0 \;\le\; \Delta F = \Delta F_0 + \sum_\ell \delta_\ell\,\Delta q_\ell \;\le\; (1 + \delta_{\max}/\beta)\,\Delta F_0 .
  \label{eq:capacity}
\end{equation}
On DeepSeek-V4-Flash, $\delta_{\max}/\beta = 0.0056/0.184 = 3.04\%$, so the indicator correction adds at most 3.04\% to $\Delta F_0$ (0.90\% on development routes under \cref{eq:capacity} from 4,864 to 5,702 experts: $\Delta F_0 = 2.219$\,ms, $\Delta F = 2.239$\,ms).

\textbf{Overlap ceiling.} Since $C + G - \max\{C, G\} = \min\{C, G\}$, concurrent branches hide at most the modeled budget $U = \sum_\ell \Ex[\min\{C(r_\ell), G_\ell\}]$ per step. Over 160 decode steps on an earlier 4,864-slot build that ran CPU jobs on every layer ($\gamma_0 = 0.120$\,ms), overlap cuts mean step time from 38.415 to 33.538\,ms, realizing 4.877\,ms (86.9\%) of $U = 5.615$\,ms.

\begin{proposition}[Single switch between prefill paths]
\label{prop:switch}
Let prefill routes be strictly stationary across token positions, let $\bar u(n)$ denote the expected distinct cold experts selected by the first $n$ tokens, and let the expected cost difference between the split and pure-upload paths take the form $\Phi(n) = \phi_0 + \phi_1 n - \phi_2\,\bar u(n)$ with $\phi_0 < 0$, $\phi_1 > 0$, and $\phi_2 > 0$. Then there exists $n^*$ such that split execution is cheaper for every $n < n^*$ and uploading for every $n > n^*$.
\end{proposition}

\begin{proof}
Let $E_{e,i}$ be the event that token position $i$ selects cold expert $e$ among $E_c$ cold experts. Then
\begin{equation*}
  \bar u(n+1) - \bar u(n) = \sum_e \Pr\bigl(E_{e,n+1} \cap \bigcap_{i=1}^{n} E_{e,i}^{c}\bigr) \ge 0 .
\end{equation*}
Dropping $E_{e,1}^c$ and applying stationarity gives
\begin{equation*}
  \bar u(n+2) - \bar u(n+1) \le \sum_e \Pr\bigl(E_{e,n+2} \cap \bigcap_{i=2}^{n+1} E_{e,i}^{c}\bigr) = \bar u(n+1) - \bar u(n),
\end{equation*}
so $\bar u(n)$ is nondecreasing, discretely concave, and bounded above by $E_c$. Consequently, $\Phi(n+1) - \Phi(n) = \phi_1 - \phi_2\bigl(\bar u(n+1) - \bar u(n)\bigr)$ is nondecreasing in $n$, while $\Phi(0) = \phi_0 < 0$ and $\Phi(n) \ge \phi_0 + \phi_1 n - \phi_2 E_c \to \infty$. Defining $n^* = \min\{n : \Phi(n) \ge 0\} \ge 1$ gives $\Phi(n) < 0$ for $n < n^*$ and, since $\Phi(n^*) - \Phi(n^* - 1) > 0$ bounds all subsequent increments below, $\Phi(n) > 0$ for $n > n^*$.
\end{proof}

Summing over layers preserves this form for the pure-upload path on GLM-5.3-Flash and Qwen3.8-Flash-Next (DeepSeek-V4-Flash calibrates $N_p = 512$ empirically alongside its top-$k_u$ selector), with per-layer costs
\begin{align*}
  \Pi^{\text{split}}(n) &= \sigma_s(n) + C_{\mathrm{pre}}\bigl(r(n),\,u(n)\bigr) + G_h(n) - \Omega_s(n),\\
  \Pi^{\text{up}}(n) &= \sigma_u(n) + \bigl(w\,u(n) + \xi(n)\bigr)/B + G_h(n) + G_c(n) - \Omega(n),
\end{align*}
where $\sigma_s, \sigma_u$ are unoverlapped setup and joins, $C_{\mathrm{pre}}\bigl(r(n),\,u(n)\bigr)$ counts one host weight read per distinct cold expert plus CPU compute across $r(n)$ cold routes, $G_h, G_c$ are GPU compute on resident and uploaded experts, $\xi$ is transfer padding, and $\Omega_s, \Omega$ are CPU--GPU and copy--compute overlap (staging gathers hidden behind DMA). Linear dependence on $n$ and $u(n)$ gives $\Ex[\Pi^{\text{split}}(n) - \Pi^{\text{up}}(n)] = \phi_0 + \phi_1 n - \phi_2\,\bar u(n)$, with $\phi_2 = w/B$ minus one host read (0.38\,ms on DeepSeek-V4-Flash).

\textbf{Split execution versus whole-layer placement.} Let a short pass touch $D_\ell$ expert bytes and $R_\ell$ routes in layer $\ell$, with \sys holding shares $h_\ell, p_\ell$ and whole-layer placement holding $h^L_\ell, p^L_\ell$. When each split layer's GPU branch finishes inside its CPU branch and the CPU pays $\theta_{B\ell}$ per byte read and $\theta_{R\ell}$ per route computed, \cref{eq:pass} yields \cref{eq:phase}, collecting split setup, GPU work, and return overhead in $\sigma$ and overlap differences in $\omega$. Because the two weighted terms separately credit saved byte reads and saved route evaluations, split execution can win with fewer touched bytes under routing skew: at the DeepSeek-V4-Flash crossover (\cref{app:baselines}), the pruned pool holds 29.6--33.5\% of touched bytes versus 50.9\%, yet serves 55.7--57.7\% of routes versus 48.8\%.

\textbf{Calibration.} Branch timings ($\gamma_0 = 0.120$\,ms, $\alpha = 0.140$\,ms, $\beta = 0.184$\,ms, $G_\ell = 0.135$--$0.139$\,ms) were calibrated in-engine on the 4,864-slot DeepSeek-V4-Flash development build; the device-side decode graph (\cref{sec:design-decode}) omits CPU requests on $r_\ell = 0$ layers ($\gamma_0 = 0$, leaving $g_\ell = G_\ell$). Per-layer CPU latency across development turns is slightly sublinear for $r \ge 1$ (0.188\,ms marginal at $r = 2$ to 0.150\,ms at $r = 6$), with $\alpha + \beta r$ within 0.01\,ms below empirical means for $r \le 3$ (98\% of held-out layer steps) and 0.077\,ms above rare $r = 6$ steps. On Qwen3-235B-A22B (per-segment graphs), calibrating its 48 layers of 10.6\,MB experts and 46 layers of 12.2\,MB experts gives $\beta = 0.139, 0.157$\,ms, $\alpha = 0.128, 0.129$\,ms, $\gamma_0 = 0.118, 0.119$\,ms, and mean $G_\ell = 0.113$\,ms ($g_\ell = \gamma_0$, $\alpha - \gamma_0 \approx 0.01$\,ms).

\end{document}